\documentclass[12pt]{iopart}
\usepackage{lipsum}
\usepackage{bm}
\usepackage{verbatim}
\usepackage{color} 
\usepackage{graphicx}
\usepackage{amssymb}
\usepackage{amsthm}

\usepackage{ragged2e}
\usepackage{hyperref}

\usepackage{lineno}
\usepackage{mathtools}
\DeclarePairedDelimiter\floor{\lfloor}{\rfloor}

\usepackage[linesnumbered,ruled,vlined]{algorithm2e}
\SetKwInput{KwInput}{Input}                % Set the Input
\SetKwInput{KwOutput}{Output}              % set the Output

\DeclareMathOperator*{\argmin}{arg\,min} % thin space, limits underneath in displays
\newtheorem{theorem}{Theorem}
\newtheorem{lemma}{Lemma}
\newtheorem{propo}{Proposition}
\newtheorem{coro}{Corollary}
\newtheorem{definition}{Definition}

\newtheorem{remark}{Remark}
\newtheorem{hypothesis}{Hypothesis}

\begin{document}

\title[Robust reconstruction of sparse network dynamics]{Robust reconstruction of sparse network dynamics}

% Use letters for affiliations, numbers to show equal authorship (if applicable) and to indicate the corresponding author
\author{Tiago Pereira$^{\dagger,\ddagger}$, Edmilson Roque dos Santos$^\dagger$ and Sebastian van Strien$^\ddagger$}
\address{$^{\dagger}$Instituto de Ci\^encias Matem\'aticas e Computa\c{c}\~ao, Universidade de S\~ao Paulo, S\~ao Carlos, Brazil}
\address{$^{\ddagger}$Department of Mathematics, Imperial College London, London, SW7 2AZ, UK}
\ead{edmilson.roque.usp@gmail.com}

\begin{abstract}
Reconstruction of the network interaction structure from multivariate time series is an important problem in multiple fields of science. This problem is ill-posed for large networks leading to the reconstruction of false interactions. We put forward the Ergodic Basis Pursuit (EBP) method that uses the network dynamics' statistical properties to ensure the exact reconstruction of sparse networks when a minimum length of time series is attained. We show that this minimum time series length scales quadratically with the node degree being probed and logarithmic with the network size. Our approach is robust against noise and allows us to treat the noise level as a parameter. We show the reconstruction power of the EBP in experimental multivariate time series from optoelectronic networks.
\end{abstract}

%\maketitle
\maketitle

%\linenumbers

\tableofcontents

\section{Introduction}

Networks of coupled dynamical systems are successful models in diverse fields of science ranging from biology \cite{winfree2001}, chemistry \cite{Kuramoto84} to physics \cite{RMP} and neuroscience \cite{zamora2011}. The network interaction structure impacts the dynamics \cite{sporns2005human,Tanzi_2020}, in fact, many malfunctions are associated with disorders in the network structure \cite{bohland2009}. While we cannot measure the network structure, we have access to a multivariate time series of nodes' states. Thus, reconstructing the network structure from multivariate data has attracted much attention by merging dynamical systems techniques \cite{takens2006detecting,Guckenheimer_2014} and optimization \cite{casadiego2017model,wang2018,Eroglu_PRX_2020}.

When the network is moderately large, the amount of data required for successful reconstruction is too large, making the network reconstruction from data a non-trivial task. Indeed, in general, the network reconstruction becomes ill-posed and unstable \cite{Domenico_2008,NOVAES_2021_basis_adaptation}. Recent strategies aim at incorporating the sparsity in the network interaction structure to formulate a minimization problem that searches sparse representations of the input data \cite{Domenico_2008,Chang2014,han2015robust,mangan2016inferring,wang2016,Pan_2016}. The key idea is that sparsity may promote a decrease in the data length required for the reconstruction \cite{Tran_2017,Schaeffer_2018}. Nevertheless, ensuring exact network reconstruction, even in the presence of sparse interactions, remains an important open problem. 

Here, we put forward the \emph{Ergodic Basis Pursuit} (EBP) method that reconstructs sparse networks from a limited amount of data.  Our method adapts the search for sparse solutions to the statistical properties of the network. We formulate the EBP as a Basis Pursuit problem adapted to the ergodicity of the network.  When the network dynamics is ergodic and has decay of correlations, we show that the reconstruction is exact once a minimal length of the time series $n_0$ is attained. We show that $n_0$ scales quadratically with the node degree and $\log$ of the system size. We also show that the reconstruction is robust against random perturbations. We illustrate the applicability of our method in experimental optoelectronic networks. Our approach enables us to treat the noise level as a tuning parameter to identify the network structure robustly. 

\section{Dynamics on complex networks} We consider dynamics as
\begin{linenomath}
\begin{align}\label{eq:net_dynamics}
%\begin{split}
x_i(t + 1) = f_i(x_i(t)) + \alpha \sum_{j = 1}^N A_{ij} h_{ij}(x_i(t),x_j(t)),
%\end{split}
\end{align}
\end{linenomath}
for each $i \in [N] := \{1, \dots, N\}$, where $x_i$ represents the state of node $i$, $f_i \colon M_i\to M_i$ corresponds to the isolated map over a bounded set $M_i \subset \mathbb{R}$, $\alpha$ is the  coupling strength, $A_{ij}$ equals $1$ if node $i$ receives a connection from $j$ and $0$ otherwise, and $h_{ij} \colon M_i \times M_j \to M_i$ is the pairwise coupling function. We denote the state of the full network as $x = (x_1, \dots, x_N)\in M^N \equiv \prod_{i \in [N]}M_i$, and $x(t + 1) = F(x(t))$. This class of networks is common in applications such as laser dynamics \cite{hart2019topological} and can be generalized to higher dimensions. 
We consider four assumptions on the network dynamics in  \eqref{eq:net_dynamics}.

\noindent
\label{assumption:net_lib} \emph{(i) Network library.} The isolated maps $f_i$ and the coupling functions $h_{ij}$ lie in the span of an ordered library $\mathcal{L} = \{\phi_1, \phi_2, \dots, \phi_m\}$ where $\phi_l \colon M^N \to \mathbb{R}$. We consider the polynomials of two variables with degree at most $r$ 
\begin{linenomath}
\begin{align}\label{eq:homogeneous_polynomials}
%\begin{split}
\mathcal{L} = \{1\} \cup \{x_i^p\}_{i, p} \cup \{x_i^p x_j^q\}_{i,j, p, q}, 
%\end{split}
\end{align} 
\end{linenomath}
where $i, j \in [N]$ with $i \neq j$ and we remove any redundancy, $p \in [r], q \in [r - 1],$ and $p + q \leq r$. The cardinality of $\mathcal{L}$ is given by $m = \binom{N}{2} \binom{r}{2} + N r + 1$. A directed edge from $j$ to $i$ is given by the presence of a nonzero coefficient in $x_i^p x_j^q$. We discuss the ordering of $\mathcal{L}$  in Section \ref{sec:net_lib}. We say that an ordered library $\mathcal{L}$ is a network library and the functions depend only on pairs of coordinates. Thus, nonzero coefficients in the network representation in $\mathcal{L}$ can be identified with links in the network structure.

\noindent
\label{assumption:sparse_net} \emph{(ii) Sparse network.} We assume that the network structure is directed and sparse, that is, for each node $i$, $x_i(t + 1) = \sum_{l=1}^{m} c_i^l \phi_l(x(t))$ for all $t \geq 0$, where $c_i = (c_i^1, \dots, c_i^m) \in \mathbb{R}^{m}$ is an $s$-sparse vector, that is, at most $s$ of its entries are nonzero, see definition \ref{def:sparse_vec}. Notice that a node $i$ with degree $k_i$ will have a number of nonzero entries of $c_i$ growing linearly with $k_i$.

\noindent
\label{assumption:exp_decay} \emph{(iii) Exponential mixing.} We assume $(F, \mu)$ satisfies exponential mixing conditions \cite{Hang_2017_bernstein-type} for the  physical measure $\mu$: given a constant $\gamma > 0$ for all $\psi \in \mathcal{C}^1(M^N;M)$ and $\mu$-integrable function $\varphi$, there exists $K(\psi, \varphi) > 0$ such that for any $t \geq 0$
\begin{linenomath}
\begin{align}
%\begin{split}
\Big|\int\psi \cdot (\varphi \circ F^t) d\mu
- \int \psi d\mu \int \varphi d\mu \Big| \leq K(\psi, \varphi)e^{-\gamma t}.
%\end{split}
\end{align}
\end{linenomath}

This assumption is typical for chaotic dynamical systems. 

\noindent
\label{assumption:prod_measure} \emph{(iv) Near product structure.} Since we are dealing with pairwise interactions, given a small $\zeta > 0$ we assume that the network physical measure $\mu$ is close to a product measure $\nu$, i.e.,  $d(\mu, \nu) < \zeta$, where $d$ calculates the maximum difference between integrals with respect to $\mu$ and $\nu$ over pair of functions in a suitable network library, see Section \ref{ssec:metric_prob_measures} for the formal definition. In the weak coupling regime, this assumption is fulfilled \cite{Eroglu_PRX_2020}. However, this assumption also holds in other scenarios, as we will illustrate later in an experimental application.

First, we consider the network reconstruction problem for the noiseless case and establish the minimum length of time series $n_0$ so that the EBP exactly reconstructs the network structure. Next, we show that the reconstruction is robust against additive measurement noise, which opens the possibility to apply for the experimental setting.

\noindent
\section{Reconstruction problem} To reconstruct the network structure $A$ from the multivariate time series data $\{x(t)\}_{t \geq 0}^n$ of   \eqref{eq:net_dynamics}, we consider the library matrix
\begin{linenomath}
\begin{align}
\Phi(X) =
\frac{1}{\sqrt{n}}
\left(
\begin{array}{ccc}
\phi_1(x(0)) 	    & \cdots & \phi_m(x(0)) \\
\phi_1(x(1)) 	    & \cdots & \phi_m(x(1)) \\
\vdots &  \ddots & \vdots \\ 
\phi_1(x(n-1)) 	    & \cdots & \phi_m(x(n-1)) \\
\end{array}
\right)
\end{align}
\end{linenomath}
and arrange the trajectories into a matrix
\begin{linenomath} 
\begin{align}\label{eq:matrix_true_time_series}
\bar{X} = \left(
\begin{array}{ccc}
x_1(1) & \cdots &  x_N(1)	    \\
\vdots    &\ddots & \vdots\\
x_1(n)  &   \cdots &  x_N(n)	  
\end{array} \right).
\end{align}
\end{linenomath}
\noindent
We aim to find the $m \times N$ matrix of coefficients $C$, which has column vectors given by $\{c_1, c_2, \dots, c_N\} \subset \mathbb{R}^m$, such that 
\begin{linenomath}
\begin{align}\label{eq:linear__C}
 \bar{X} = \Phi(X) C.   
\end{align}
\end{linenomath}
When the amount of data is large in comparison to the network size, the library matrix $\Phi(X)$ might be full column rank and (sparse) approximations can be found by least square based-methods \cite{brunton2016discovering,mangan2016inferring,wang2018}. For short time series, the matrix $\Phi(X)$ has more columns than rows, and one approach is to solve for each node $i$ the basis pursuit (BP) problem
\begin{linenomath}
\begin{align}\label{eq:BP}
(\mathrm{BP}) \quad \min_{u \in \mathbb{R}^m} \| u \|_{1} \quad \mbox{subject to} ~\Phi(X) u = \bar{x}_i,
\end{align}
\end{linenomath}
where $\bar{x}_i$ is the $i$-th column $\bar{X}$. This implementation was used for networks of moderate size \cite{wang2016,Schaeffer_2018,Schaeffer_2020}. For large networks, this may lead to spurious linear dependencies among the columns $\Phi(X)$ \cite{Domenico_2008,NOVAES_2021_basis_adaptation}, and  \eqref{eq:BP} does not have a unique sparse solution. In Figure \ref{fig:basis_comparison} b) and c), we show that the basis pursuit (in purple) requires a minimum length of time series $n_0$ that scales with the system size to reconstruct a ring network in coupled logistic maps, although the network is sparse. The BP method is inappropriate for large-scale networks.

\section{Main results: informal statements}

To establish conditions for the uniqueness of the reconstruction, successful approaches show that any set of $2s$ columns of $\Phi(X)$ is nearly orthonormal, what is known as the restricted isometry property (RIP) \cite{Candes_2005_decoding}. By noticing that the inner product of pair of columns of the original library matrix $\Phi(X)$ can be represented as a Birkhoff sum. We introduce a new library $\mathcal{L}_{\nu}$ by a Gram-Schmidt process such that the new columns have vanishing Birkhoff sum when the length of the times series diverges. Then,  we use a concentration inequality to estimate the minimal length of the time series such that any two distinct column vectors of $\Phi_{\nu}(X)$ are nearly orthonormal to each other, and RIP for the desired sparsity.

This strategy is implemented with four main results: 
(I) the introduction of the new basis $\mathcal{L}_{\nu}$ is a network library which keeps the appropriated sparsity of the original problem; (II) It is possible to obtain a desired RIP constant for $\Phi_{\nu}(X)$ by exploring the ergodicity of the dynamics; and  (III) the reconstruction is unique.  Finally, we show (IV) the robustness against measurement noise.

\noindent
\subsection{Constructing the adapted network library} {\color{black} First, we notice that using the invariant measure $\mu$ directly to obtain a new basis and thus an almost orthonormal structure in the columns of the corresponding $\Phi(X)$ leads to the loss of sparsity in the representation. Indeed, the new orthonormal basis would contain functions that depend on all coordinates, e.g., of the form $\varphi(x_1, \dots, x_N)$ because $\mu$ is not a product measure and mixes all coordinates. Hence, we consider a product measure $\nu$ close to $\mu$.   }More precisely, we perform a Gram-Schmidt (GS) process in the span of $\mathcal{L}$ and obtain a basis $\hat{\mathcal{L}} = \{ \hat \varphi_1, \dots, \hat \varphi_m\}.$ We define $ \varphi_i = a_i  \hat \varphi_i$, where $a_i^2 = 1/\int \hat \varphi_i^2 d\nu$, so the new basis $\mathcal{L}_{\nu} = \{ \varphi_i \}_{i=1}^{m}$ is an orthonormal system with respect to a product measure $\nu$. We assume that each marginal of $\nu$ is absolutely continuous with respect to Lebesgue, and the corresponding density is Lipschitz. For our network library $\mathcal{L}$ we have that:
\begin{itemize}
\item[] \textbf{Theorem \ref{thm:net_lib_preserved} (Network library is preserved).} GS process maps an $s$-sparse representation of $F$ in $\mathcal{L}$ to an $\omega_r(s)$-sparse representation in the orthonormal network library $\mathcal{L}_{\nu}$.
\end{itemize}
Here,   $\omega_r(s) = \Big( \floor*{\frac{r}{2}} \big(r - \floor*{\frac{r}{2}} \big)  + r + 1\Big) s$, where $\floor*{\beta}$ denotes the largest integer $p$ satisfying $p \leq \beta$.
We call $\mathcal{L}_{\nu}$ the adapted network library and denote the respective library matrix by $\Phi_{\nu}(X) = \Phi(\mathcal{L}_{\nu}, X)$.  The proof uses that the GS process is a recursive method that involves projections onto preceding functions. Since $\nu$ is a  product, the projections of the GS are split into products of integrals. Thus, $\mathcal{L}_{\nu}$ does not have functions that depend on more than two variables and characterize a network library. Also, the representation remains sparse, see Section \ref{sec:net_is_preserved} for details.

\noindent
\subsection{Ergodic basis pursuit} Notice that since library $\mathcal{L}_{\nu}$ is orthonormal, the set of columns vectors of $\Phi_{\nu}(X)$ form a set of $s$ nearly orthonormal column vectors. We quantify the orthonormality via the $s$-th restricted isometry constant $\delta_s = \delta_s(\Phi_{\nu}(X))$ as the smallest $\delta \geq 0$ such that
\begin{linenomath}
\begin{align}\label{eq:RIP_property}
(1 - \delta)\|u\|_2^2 \leq \|\Phi_{\nu}(X) u\|_2^2 \leq (1 + \delta)\|u\|_2^2
\end{align}
\end{linenomath}
for all $s$-sparse vectors $u \in \mathbb{R}^m$. 
%We say that the matrix $\Phi_{\nu}(X)$ satisfies the restricted isometry property (RIP) when $\delta_s \in (0, 1)$. 
Next, we determine the minimum length of time series such that $\Phi_{\nu}(X)$ is RIP with a desired small $\delta_s$. Our second result is
\vspace{-0.1cm}
\begin{itemize}
\item[] \textbf{Theorem \ref{thm:appendix_noiseless_case}.\ref{thm:phi_nu_rip} ($\Phi_{\nu}(X)$ satisfies RIP).} Consider  $d(\mu, \nu) < \zeta$ for sufficiently small $\zeta$. For large network sizes and large set of initial conditions if the length of time series $n$ is at least
\begin{linenomath}
\begin{align}\label{eq:approx_min_lgth_time_series}
n_0 \approx K_1 \omega_r(s)^2 \ln{(N r)},
\end{align}
\end{linenomath}
for a positive constant $K_1$, then $\Phi_{\nu}(X)$ satisfies  \eqref{eq:RIP_property} with $\delta_{2\omega_r(s)} \leq \sqrt{2} - 1$.
\end{itemize}
\vspace{-0.1cm}
{\color{black}The proof is presented in Section \ref{sec:EBP_has_unique_sol} and the key steps are as follows. 
First, 
we use the coherence \cite{Donoho_uncertanty_principle,Donoho_2006_mutual_coherence} defined as
\begin{linenomath}
\begin{align*}
\eta (\Phi_{\nu}):= \max_{i \neq j} \left| \langle v_i, v_j \rangle  \right|
\end{align*} 
\end{linenomath}
over distinct pairs of normalized (Euclidean norm) columns of $\Phi_{\nu}(X)$. Since we know that $\delta_s \leq \eta(\Phi_{\nu}) (s-1)$ for any  $s \geq 2$ \cite{Foucart_mathematical_compressive_sensing}, it suffices to introduce a library $\Phi_{\nu}(X)$ whose coherence is small enough to obtain the desired RIP constant.

Second, let $v_i$ be the $i$-th column of the matrix $\Phi_{\nu}(X)$ and notice that the inner product between columns $i$ and $j$ is
\begin{linenomath}
\begin{align*}%\label{eq:inner_product_u}
\langle v_i, v_j \rangle = 
\frac{1}{n} \sum_{t=0}^{n - 1} (\varphi_i \cdot  \varphi_j )\circ (F^t(x(0))).
\end{align*} 
\end{linenomath}
Using that $\mu$ and $\nu$ are close, by triangular inequality and the Bernstein-type inequality, see \cite{Hang_2017_bernstein-type}, we control the coherence $\eta (\Phi_{\nu})$ by approximating it by $\int \varphi_i \cdot  \varphi_j  d \mu$. Hence, we can determine a large set of initial conditions such that the RIP of $\Phi_\nu(X)$ is less than $\sqrt{2}-1$, see Section \ref{sec:proof_n_0}. Since $\Phi_\nu(X)$ is RIP, we obtain}
\begin{itemize}
\item[] \textbf{Theorem \ref{thm:appendix_noiseless_case}.\ref{thm:ebp_has_unique_sol} (EBP has unique solution).} The convex problem that we call Ergodic Basis Pursuit
\begin{linenomath}
\begin{align}\label{eq:main_ergodic_BP}
%\begin{split}
\mathrm{(EBP)} \quad  \min_{u \in \mathbb{R}^{m}} \|u\|_1 \quad \mbox{subject to} ~\Phi_{\nu}(X) u = \bar{x},
%\end{split}
\end{align}
\end{linenomath}
has a unique $\omega_r(s)$-sparse solution. That is, $c_{\nu}$ is the only solution of this minimization problem when $\bar{x} = \Phi_{\nu}(X)c_{\nu}$. 
\end{itemize}
The proof follows from Theorem \ref{thm:appendix_noiseless_case}.\ref{thm:phi_nu_rip}. EBP can be applied as a network reconstruction method. In terms of the coefficients $\{c_1, \dots, c_N\}$, we create a weighted edge between node $i$ and $j$ using
\begin{linenomath}
\begin{align}\label{eq:weighted_edge}
W_{ij} = \max_{k \in \mathcal{S}_j} c_i^k.
\end{align}
\end{linenomath}

We reconstruct a weighted subgraph using the node $i$, its neighbors, and the entry's magnitude of $c_i$ as the edge weight; see details in the SI \ref{sec:net_selection}.

\begin{remark}[Minimum length of time series for networks]
The degree distribution and the condition in  \eqref{eq:approx_min_lgth_time_series} can be used to estimate the amount of data that ensures the network reconstruction.
\begin{itemize}
\item \textbf{Erd\H{o}s-R\'{e}nyi (ER) networks.} We can apply \eqref{eq:approx_min_lgth_time_series} in $\mathcal{O}(1)$ for known random networks. First, note that $\omega_r(s)$ is a linear function with the sparsity level $s$, and consequently, it is a linear function of the degree $\propto k_i$ of the node $i$. Also, $\omega_r(s) < (r + 1)^2$. The degree distribution is given by a Poisson distribution, so by concentration inequality \cite{FanChung}, most nodes have their degree close to the mean degree $\langle k \rangle$. Hence, to reconstruct a typical node in ER networks requires (in $\mathcal{O}(1)$) the minimum length of time series given by
\begin{linenomath} 
\begin{align}\label{eq:er_scaling}
n_0 = \mathcal{O}\big((r + 1)^4 \langle k \rangle^2 \ln{(N r})\big).
\end{align} 
\end{linenomath} 
Note that  $\langle k \rangle = p N$, where $p$ is the probability of including an edge in the graph. In the phase where ER network becomes almost sure connected, $p = K \ln{N}/N$ with $K \geq 1$ \cite{FanChung}. So,
\begin{linenomath} 
\begin{align*}
n_0 = \mathcal{O}\big((r + 1)^4 \ln(N)  \ln{(N r})\big).
\end{align*} 
\end{linenomath} 
\item \textbf{Scale-free networks.} In scale-free networks, the same growth scaling \eqref{eq:er_scaling} is valid for low-degree nodes. However, hubs in Barabási-Albert networks have their degree proportional to $N^{\frac{1}{2}}$, so it requires
\begin{linenomath} 
\begin{align*}
n_0 = \mathcal{O}\big((r + 1)^4\langle k \rangle^2 N \ln{(N r)}\big).
\end{align*}
\end{linenomath} 
\item \textbf{Regular networks.} All nodes have the same degree. So, the same growth scaling \eqref{eq:er_scaling} is valid for any node in the network.
\end{itemize}
\end{remark}
%For Erd\H{o}s-R\'{e}nyi (ER) networks $n_0 = \mathcal{O}\big((r + 1)^4 \langle k \rangle^2 \ln{N r}\big)$.  scale-free networks, as opposed to hubs in Barabási-Albert networks, requiring $n_0 = \mathcal{O}\big((r + 1)^4\langle k \rangle^2 N \ln{(N r)}\big)$, see Section \ref{sec:rand_nets} for details.

%\section{}
%\label{sec:rand_nets}

\noindent
\subsection{Robust reconstruction} We now extend the EBP to measurements corrupted by noise
\begin{linenomath}
\begin{align}\label{eq:noisy_observation}
    y(t) = x(t) + z(t), 
\end{align}
\end{linenomath}
where $(z_n)_{n \geq 0}$ corresponds to independent and identically distributed $[-\xi, \xi]^{N}$-valued noise process, with probability measure $\rho_{\xi}$. The probability measure of the process $(y_n)_{n \geq 0}$ is the convolution $\mu_{\xi} := \mu * \rho_{\xi}$ \cite{folland2013real}, which converges weakly to $\mu$ as $\xi \to 0$. We assume that $\mu_{\xi}$ is estimated using a product measure $\nu$.   We use that $\mu_{\xi}$ is close to $\nu$ to estimate a new bound for the minimum length of the time series $\tilde{n}_{0}$ such that $\Phi_{\nu}(X)$ satisfies RIP with constant $\delta_{2\omega_r(s)} \leq \sqrt{2} - 1$.%, as stated in SI \ref{sec:noise_robust_reconstruction}. 

Since we measure the corrupted data $Y$ instead of $X$, we use the Mean Value theorem to deduce that
\begin{linenomath}
\begin{align}\label{eq:phi_0_y}
    \Phi_{\nu}(Y) = \Phi_{\nu}(X) + \Lambda(X, \bar{Z}),
\end{align}
\end{linenomath}
where $\|\Lambda(X, \bar{Z})\|_{\infty} \leq m N r^2 K_1 \xi $ and $K_1 $ depends on the density of the marginals of $\nu$. The noisy observation in  \eqref{eq:noisy_observation} can be recast as a perturbed version of the orthonormal version of  \eqref{eq:linear__C} column-wise
\begin{linenomath}
\begin{align}\label{eq:perturbed_noise_}
\bar{y} = \Phi_{\nu}(Y) c_{\nu} + \bar{u},
\end{align}
\end{linenomath}
where $c_{\nu}$ is the coefficient vector associated to the network library $\mathcal{L}_{\nu}$ and $\bar{u}$ is $\ell_2$ bounded, see Section \ref{sec:noise_robust_reconstruction}. Thus, we can state our final result:
\vspace{-0.1cm}
\begin{itemize}
\item[] \textbf{Theorem \ref{thm:appendix_noise_reconstr} (EBP is robust).} If the length of time series $n \geq \tilde{n}_{0}$, then the family of solutions $\{c_{\nu}^{\star}(\epsilon)\}_{\epsilon > 0}$ to the convex problem (which we call the Quadractically constrained Ergodic Basis Pursuit)
\begin{linenomath}
\begin{align}\label{eq:QBP_EBP}
%\begin{split}
\mathrm{(QEBP)} \quad \min_{\tilde{u} \in \mathbb{R}^{m}} \|\tilde{u}\|_1 ~\mathrm{subject~to}~ \|\Phi_{\nu}(Y) \tilde{u} - \bar{y}\|_2 \leq  \epsilon
%\end{split}
\end{align}
\end{linenomath}
satisfies
\begin{linenomath}
\begin{align}\label{eq:ell_2_close_solution}
    \|c_{\nu}^{\star}(\epsilon) - c_{\nu}\|_2 \leq K_2 \epsilon
\end{align}
\end{linenomath}
for some $K_2>0$ as long as 
$
\epsilon \geq \sqrt{n} \xi \Big(1 + m N r^2 K_1 \|c_{\nu}\|_{\infty} \Big).
$
\end{itemize}

\begin{figure}[t]
    \centering
    \includegraphics[width=1.0\textwidth]{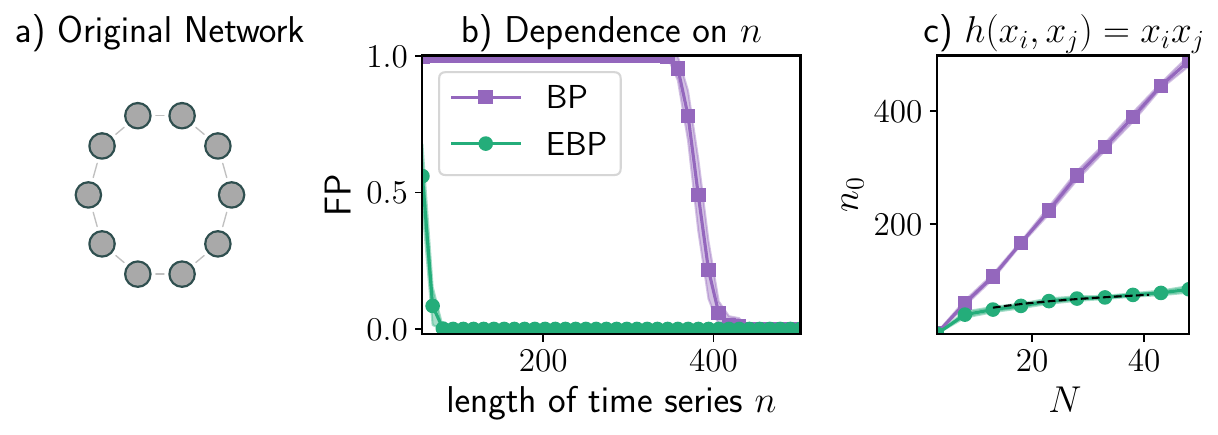}
    \caption{\textbf{Ergodic Basis Pursuit performance requires only short time series.} a) Illustration of a ring graph with $N = 10$. b) False positive (FP) of the reconstructed ring network with respect to the length of time series $n$ for a network size $N = 40$. c) The minimum length of time series $n_0$ for a successful reconstruction versus system size $N$. Basis pursuit (BP) and ergodic basis pursuit (EBP) are shown in (purple) squares and (green) circles, respectively. The network dynamics parameters are $a = 3.990$ and coupling strength $\alpha = 5 \times 10^{-4}$. The shaded area corresponds to the standard deviation with respect to 10 distinct initial conditions uniformly drawn in $[0, 1]^{N}$. The (black) dashed is the scaling $\ln N$ for reference. The Kernel density estimation of $\nu$ is used with bandwidth $\chi = 0.05$. The multivariate time series is generated without noise.}
    \label{fig:basis_comparison}
\end{figure}

\subsection{Numerical experiment: coupled logistic maps} 

To compare the reconstruction performance of the EBP against the classical BP, we consider coupled logistic maps, $f(x_i) = a x_i(1 - x_i)$ with $a = 3.990$, via the pairwise coupling function $h(x_i, x_j) = x_i x_j$ with overall coupling strength $\alpha = 5 \times 10^{-4}$. Figure~\ref{fig:basis_comparison} a) illustrates a ring network with $N=10$ nodes. 

In Figure \ref{fig:basis_comparison} b), we evaluate the reconstruction performance employing the basis pursuit (BP) and the ergodic basis pursuit (EBP) as we increase the length of time series $n$. The convex minimization problem is solved employing the \texttt{CVXPY} package \cite{diamond2016cvxpy,agrawal2018rewriting}, in particular, \texttt{ECOS} solver \cite{ECOS}. We consider the network library in \eqref{eq:homogeneous_polynomials} with degree at most $3$, so by construction, there exists a sparse representation of the network dynamics in this library. We observe that the false positive fraction -- calculates the presence of mistakenly found edges regardless of their weights -- of the BP goes to zero when $n_0 \approx 400$, roughly tenfold the system size. On the other hand, EBP outperforms the basis pursuit method, reducing the necessary length of time series to reconstruct the network. To evaluate the scaling with respect to the system size, we calculate $n_0$ as we increase $N$. In Figure c), we confirm that $n_0$ scales with the system size for BP instead of $\ln N$ of the EBP method. In the Section \ref{sec:diff_nets}, we demonstrate that our estimates of $n_0$ predict the numerical observation when we vary the maximum degree of different network structures.

\subsubsection{Coupled logistic maps under different network structures}
\label{sec:diff_nets}

\begin{figure}
    \centering
    \includegraphics[width=0.5\textwidth]{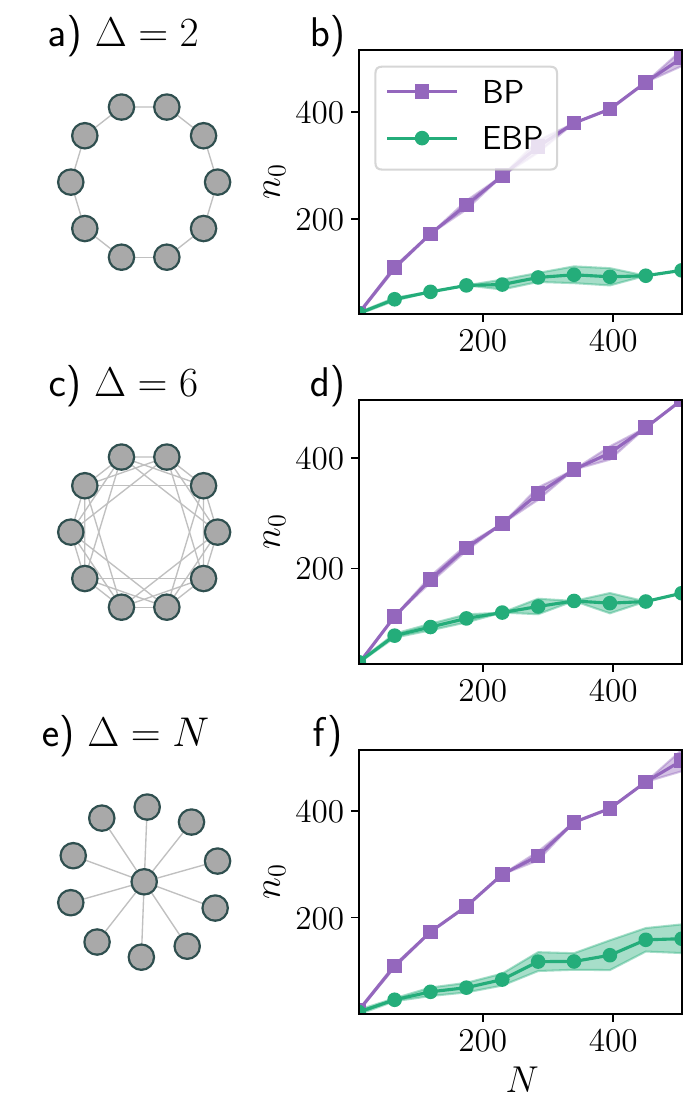}
    \caption{\textbf{Comparison between BP and EBP under different network structures.} a) Ring graph with maximum degree $\Delta = 2$. b) The minimum length of time series $n_0$ for a successful reconstruction versus system size $N$, and similarly in d) and f). c) Lattice graph with maximum degree $\Delta = 6$. e) Star graph where the maximum degree grows with the system size. Basis pursuit (BP) and ergodic basis pursuit (EBP) are shown in (purple) squares and (green) circles, respectively. The network dynamics parameters are $a = 3.990$ and coupling strength $\alpha = 1 \times 10^{-3}/\Delta$, so the coupling term in the network dynamics is normalized as we vary $N$. The shaded area corresponds to the standard deviation with respect to 10 distinct initial conditions uniformly drawn in $[0, 1]^{N}$. The Kernel density estimation of $\nu$ is used with bandwidth $\chi = 0.05$. The multivariate time series is generated without noise.}
    \label{fig:basis_comparison_under_different_nets}
\end{figure}

Here, we consider a different coupling function given by $h(x_i, x_j) = x_j^2$ and analyze for distinct network structures, see Figure \ref{fig:basis_comparison_under_different_nets}.
We observe that the EBP method outperforms the basis pursuit on all occasions. If we compare the profile of the curves, all curves look similar to each other. The difference is that in b) and d), EBP requires more data to reconstruct the network structure. This phenomenon was predicted by our estimate in the expression \eqref{eq:approx_min_lgth_time_series}. Since the maximum degree is larger, the sparsity level $s$ of the target sparse vector is also larger, implying that $n_0$ grows.

\section{Reconstruction of experimental optoelectronic networks} 
\begin{figure}
    \centering
    \includegraphics[width=1.0\linewidth]{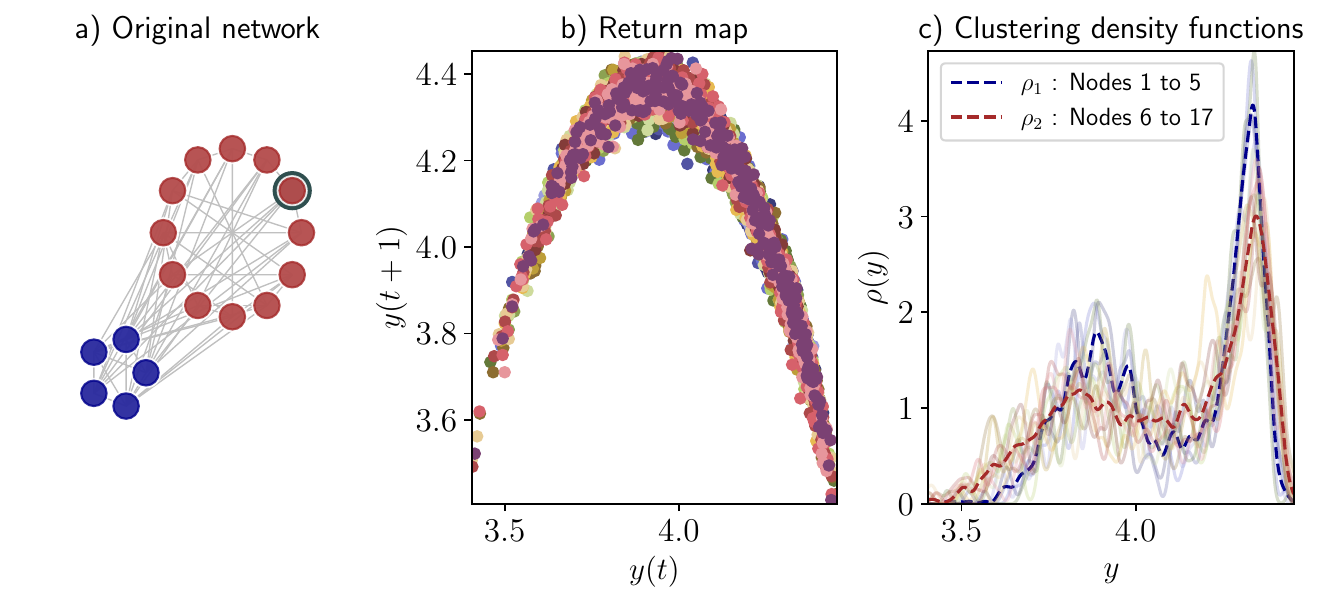}
    \caption{\textbf{Network dynamics of experimental optoelectronic data.} a) Original optoelectronic network with two groups of nodes --- dark gray node is marked for future reference. b) Return map for all nodes in the network. c) Densities function $\rho_i$ for each node $i$ (in light color) estimated using each node's time series. Clustering density estimation displays two resulting densities corresponding to two groups of nodes, in blue and red. The density estimation utilizes a Gaussian kernel with bandwidth $\chi = 0.05$.}
    \label{fig:opto_electronic_data}
\end{figure}

The data is generated from a network of optoelectronic units whose nonlinear component is a Mach-Zehnder modulator \cite{hart2019topological}. The network is modeled as
\begin{linenomath}
\begin{align}\label{eq:opto_electronic_}
x(t + 1) = \beta I_{\theta}(x_i(t)) - \alpha \sum_{j=1}^{17} L_{ij}I_{\theta}(x_j(t))  \mbox{~mod~} 2\pi,
\end{align}
\end{linenomath}
where the normalized intensity output of the Mach-Zehnder modulator is given by $I_{\theta}(x) = \sin^2(x + \theta)$, $x$ represents the normalized voltage applied to the modulator, $\beta$ is the feedback strength, $\theta$ is the operating point set to $\frac{\pi}{4}$ and $L$ is the Laplacian matrix --- $L_{ij} = \delta_{ij} k_i - A_{ij}$,
where $\delta_{ij}$ is the Kronecker delta and $k_i$ is the $i$-th node degree. The experiments were done by varying the coupling strength between the nonlinear elements in an undirected network, depicted in Figure \ref{fig:opto_electronic_data} a). We will show results for coupling $\alpha=0.171875$.

Instead of having access to trajectories from \eqref{eq:opto_electronic_}, we have access to the noisy experimental multivariate time series $\{y_{1}(t), \dots, y_{17}(t)\}_{t=1}^{264}$, whose return map is depicted in Figure~\ref{fig:opto_electronic_data} b).  Thus, we are naturally in the setting of \eqref{eq:QBP_EBP} the randomly perturbed version of the EBP.
Typically, for experimental data the noise level $\xi$ is unknown. So, we use the constraint $\epsilon$ in \eqref{eq:QBP_EBP}  as a parameter to tune and search for the correct incoming connections. 

The key idea is as follows. For large values of $\epsilon$ we have that $c_{\nu}^{\star}(\epsilon)=0$ is a solution to \eqref{eq:QBP_EBP}. Next, for moderate values of $\epsilon$, the coefficients corresponding to the isolated dynamics appear in $c_{\nu}^{\star}(\epsilon)$.  As we decrease $\epsilon$, we start observing correct connections that are present over multiple values of $\epsilon$. We aim to identify those robust connections. This can be formulated as an algorithm that we call \emph{relaxing path}, which is described in Materials and Methods. The algorithm consists in solving \eqref{eq:QBP_EBP} for multiple values of $\epsilon$ while checking which entries of $c_{\nu}^{\star}(\epsilon)$ that correspond to connections persist as $\epsilon$ varies.

To apply these ideas to the experimental data, we first perform a pre-processing. Most of the data are concentrated in a portion of the phase space with scarce excursions to other parts. Thus, we first restrict the data to a portion of the phase space mostly filled, see further details in the SI \ref{sec:opto_electronic_details}. After this procedure, we obtain a parabolic shape of the return map that corresponds to the restriction of the original optoelectronic network dynamics $F$ onto the interval $\mathcal{A} = [3.4, 4.5]$ over 264-time steps, which we denote $\tilde{F}= F|_{\mathcal{A}}$. Hence, $\tilde{F}$ lies in the span of the quadratic polynomials, and we use $\mathcal{L} = \{\phi_i^p(x_i) = x_i^p : p=0, 1, 2\}$. To perform a Gram-Schmidt process, we estimate the $\nu$ using all trajectories of a group of nodes through kernel density estimator, improving the estimate accuracy. We assume $d\nu = \rho_1 \times \rho_2 d x d y$ is a product of two densities. Nodes 1 to 5 have the density illustrated in blue, and the remaining nodes have the density in red in the right panel of Figure~\ref{fig:opto_electronic_data}. 

\begin{figure}
    \centering
    \includegraphics[width=1.0\textwidth]{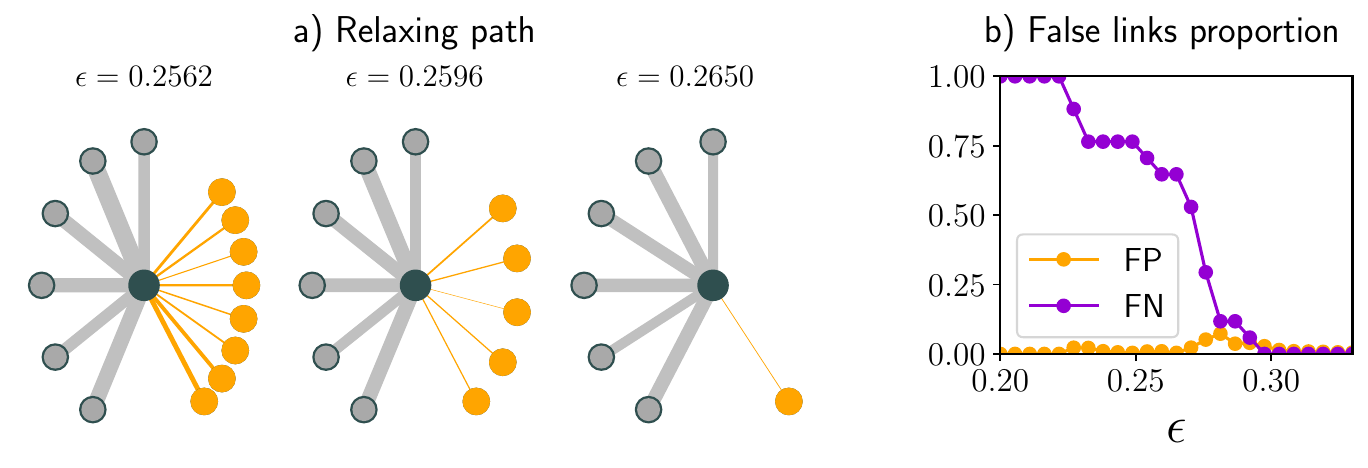}
    \caption{\textbf{Reconstruction of the original network from experimental data.} a) Relaxing path algorithm is performed in the node (in dark gray) from the left panel. There are three different relaxing parameter values, where the edges are colored accordingly: the true edges (in gray) and false positives (in orange) while the thickness is the edge weight, see  \eqref{eq:weighted_edge}. b) False positive (FP) (in orange) and false negative (FN) (in purple) of the reconstructed network versus the parameter $\epsilon$. We varied the $\epsilon$ parameter through 25 values equally spaced in the interval $\mathcal{E} = [0.20, 0.33]$. We employ ECOS convex optimization solver \cite{ECOS} to solve ~\eqref{eq:QBP_EBP}.}
    \label{fig:fig1_v0}
\end{figure}

The left panel of Figure~\ref{fig:fig1_v0} displays the relaxing path algorithm probing a node (the marked dark gray node in Figure \ref{fig:opto_electronic_data}) for three distinct $\epsilon$ values. For each $\epsilon$, we use \eqref{eq:weighted_edge} to construct  from $c_{\nu}^{\star}(\epsilon)$ the weighted subgraph corresponding to the probed node's neighbors. As we vary $\epsilon$ all edge weights decrease in magnitude (edge thickness), in particular false connections (in orange) that are not robust against variation of $\epsilon$. In fact, for the smallest $\epsilon$ (in the left) we observe a few false connections whose edge weights are smaller than the true connections (in gray). As we increase $\epsilon$, a few false connections start to vanish. Further increasing $\epsilon$ only the robust connections are present and the algorithm stops. Since the algorithm is node-dependent, we quantify the overall reconstruction performance in the parameter interval via a weighted false link proportion for each node, expressed in the SI \ref{sec:reconstr_opto}, and then average over all 17 nodes. The right panel of Figure \ref{fig:fig1_v0} shows that the algorithm identifies the original network structure successfully within an interval of the parameter $\epsilon$. 

\section{Mathematical analysis and preliminaries}
In the remainder of this paper we prove our main results Theorems \ref{thm:net_lib_preserved}, \ref{thm:appendix_noiseless_case} and \ref{thm:appendix_noise_reconstr}. To this end we briefly recall some definitions and established results from compressive sensing \cite{Foucart_mathematical_compressive_sensing} and exponentially mixing dynamical systems \cite{Hang_2017_bernstein-type}. 

We introduce the notation $[m] := \{1, 2,\dots,m\}$. For a matrix $\Phi \in \mathbb{R}^{n \times m}$ and a subset $\mathcal{S} \subseteq [m]$, $\Phi_{\mathcal{S}}$ indicates the column submatrix of $\Phi$ consisting of the columns indexed by $\mathcal{S}$. We denote the transpose of $\Phi$ as $\Phi^T$. We denote the $L^2(\mu)$ norm of a function $\psi:M^{N} \to \mathbb{R}$ as
\begin{linenomath}
\begin{align*}
    \|\psi\|_{\mu} = \Big( \int |\psi(x)|^{2} d \mu(x) \Big)^{1/2},
\end{align*}
\end{linenomath}
and denote $\|\psi\|_{\infty} := \sup_{x \in M^N} |\psi(x)|$.  Also, we denote $\floor*{\beta}$ as is the largest number $p \in \mathbb{N}$ satisfying $p \leq \beta$.

Let $\{M_i\}_{i \in [N]}$ be a collection of subsets of $\mathbb{R}$. For $\mathcal{J} \subset [N]$ denote the canonical projection by 
\begin{linenomath}
\begin{align}\label{eq:canonical_projection}
\pi_{\mathcal{J}}: M^N \to \prod_{i \in \mathcal{J}} M_i.
\end{align}
\end{linenomath}
%
%\subsection{Network dynamics}
%Here, we generalize the network dynamics equation introduced in the main text. Let $\{M_i\}_{i \in [N]}$ be a collection of subsets of $\mathbb{R}$ and $G = ([N], \mathcal{E})$ is a labelled and directed graph with adjacency matrix $A$. Let $\{f_i\}_{i \in [N]}$ be a collection of isolated maps $f_i: M_i \to M_i$, and $\{h_{i j}\}_{i, j \in [N]}$ a collection of pairwise coupling functions $h_{i j}: M_i \times M_j \to M_i$. We refer to the map $F_{\alpha}: \prod_{i = 1}^N M_i \to \prod_{i = 1}^{N} M_i$ as the network dynamics map if there exists an open subset $\mathcal{A} \subset \mathbb{R}$ such that for each coupling strength $\alpha \in \mathcal{A}$ the map is given as
%\begin{linenomath} \begin{align}\label{eq:net_dynamics_labelled}
%F_{\alpha}^i(x) = f_i(x_i) + \alpha \sum_{j = 1}^N A_{ij} h_{ij}(x_i, x_j), \quad i \in [N]. 
%\end{align}\end{linenomath}
%We denote $M^N = \prod_{i = 1}^N M_i$. 

\subsection{Network library}
\label{sec:net_lib}
%The concept of network library can be more general than the one introduced in the main text. 
Consider a network dynamics in \eqref{eq:net_dynamics}. Suppose that for each $i \in [N]$ there exists $m_i \in \mathbb{N}$ such that the isolated map $f_i$ is in the span of the set $\{\phi_{i}^{p}:p \in [m_i]\}$ of functions $\phi_{i}^{p}: M_i \to \mathbb{R}$, i.e., $f_i = \sum_{p = 1}^{m_i} c_i^p \phi_{i}^{p}$. We denote the collection of all these functions as 
\begin{linenomath} 
\begin{align*}
\mathcal{I} = \{\phi_i^p:i \in [N], p \in [m_i]\}.
\end{align*} 
\end{linenomath} 
Similarly, for each $i, j \in [N]$ there exist $m_{i}, m_j \in \mathbb{N}$ such that the pairwise coupling function $h_{ij}$ lies in the span of the set $\{\phi_{ij}^{pq}:p \in [m_{i}], q \in [m_j]\}$ of functions $\phi_{ij}^{pq}: M_i \times M_j \to \mathbb{R}$, i.e., $h_{ij} = \sum_{q = 1}^{m_j}\sum_{p = 1}^{m_{i}} c_{ij}^{pq} \phi_{ij}^{pq}$. We denote the collection of all these functions as
\begin{linenomath} 
\begin{align*}
\mathcal{P} = \{\phi_{ij}^{pq}: i,j \in [N], p \in [m_{i}], q \in [m_j]\}.
\end{align*} 
\end{linenomath} 
We remove any redundancy in the collections $\mathcal{I}$ and $\mathcal{P}$. In particular, we make explicit the constant function $1$ to avoid a trivial redundancy. We define the network library:

\begin{definition}[Network Library]\label{def:net_library}
We call \emph{network library} the collection of functions 
\begin{linenomath} \begin{align}\label{eq:net_lib}
\mathcal{L} = \{1\} \cup \mathcal{I} \cup \mathcal{P}
\end{align}\end{linenomath}
that represent the network dynamics map $F_{\alpha}$ in  \eqref{eq:net_dynamics}. 
\end{definition}

The network library can capture the network structure because the basis functions correspond to pairwise interactions. For the node $i$ dynamics, a nonzero coefficient of $\phi_{ij}^{pq} \in \mathcal{L}$ are associated with an edge between node $i$ and $j$ in the network. More precisely, the node $i$ of the network is identified by the labeled coordinate on $M_i$. The following definition identifies the edge:

\begin{definition}[Edge via Network Library]\label{def:edge_net_library}
Let $i \in [N]$ and $F_i$ has a representation in $\mathcal{L}$. Let $\mathcal{L}_i \subset \mathcal{L}$ be a subset that contains all necessary basis functions such that $F_i \in \mathrm{span}~\mathcal{L}_i$. If $\phi_{ij}^{pq} \in \mathcal{L}_i$ for $j \in [N]$, $p \in [m_i], q \in [m_j]$, then there is an directed edge from $j$ to $i$. 
\end{definition}

A priori, the network library has no natural ordering, so that we can introduce an ordered network library. We choose the following ordering: it first disposes of the constant function. Then, it is followed by the functions in $\mathcal{I}$, which are ordered fixing the $i \in [N]$ and letting run the index $p \in [m_i]$. Finally, the set $\mathcal{P}$ is ordered, fixing an element of the index set \{$(i, j) \in [N] \times [N]$\} (which is organized in lexicographic order) and running through the index set $\{(p, q) \in [m_i] \times [m_j]\}$ (also organized in lexicographic order), i.e.,
\begin{linenomath} \begin{align}\label{eq:ord_net_lib}
\begin{split}
\mathcal{L}^o &= \{1, \phi_1^1(x_1), \dots, \phi_{1}^{m_1}(x_1), \phi_2^1(x_2), \dots, \phi_2^{m_2}(x_2), \dots, \phi_N^1(x_N),\dots, \phi_N^{m_N}(x_N),\\
&\qquad \qquad \phi_{11}^{11}(x_1, x_1), \dots, \phi_{NN}^{m_{N}m_{N}}(x_N, x_N)\}.
\end{split}
\end{align}\end{linenomath}
We abuse notation and denote the ordered network library simply as $\mathcal{L}$.

We also define an $s$-sparse representation of the network dynamics $F_{\alpha}$ in a network library. Let us define an $s$-sparse vector.

\begin{definition}[Sparse vector]\label{def:sparse_vec} 
A vector $u \in \mathbb{R}^m$ is said to be $s$-sparse if it has at most $s$ nonzero entries, i.e.,
\begin{linenomath}
\begin{align*}
|\{j \in \{1, \dots, m\} : u^j \neq 0 \}| \leq s.
\end{align*} 
\end{linenomath} 
\end{definition}

Each node in the network has its sparsity level in the library, but we consider an upper bound in the sparsity level to depend only on one parameter $s$. To make notation easier in next definition, let $\mathcal{L} = \{\phi_l: M^N \to \mathbb{R}: l \in [m]\}$ be the network library, where $m$ is its cardinality. 

\begin{definition}[Sparse Network Dynamics Representation]
$F_{\alpha}:M^N \to M^N$ has an $s$-sparse representation in $\mathcal{L}$ if there exists a set $\{c_1, \dots, c_{N}\} \subset \mathbb{R}^{m}$ of $s-$sparse vectors such that the coordinate $i \in [N]$ is given by $F_i = \sum_{l = 1}^m c_i^l \phi_l$, where $c_i = (c_i^1, \dots, c_{i}^{m}) \in \mathbb{R}^m$.
\end{definition}

\subsection{Sparse recovery}
Here we outline the results of sparse recovery employed in the paper. The next proposition states an equivalent expression to the restricted isometry constant and restricted isometry property.
\begin{propo}
The $s-$th restricted isometry constant $\delta_s$ is given by 
\begin{linenomath} 
\begin{align*}
\delta_s = \max_{\mathcal{S} \subset [m], \mathrm{card}(\mathcal{S}) \leq s} \|\Phi_{\mathcal{S}}^{T}\Phi_{\mathcal{S}} - \mathbf{1}_{s}\|_2,
\end{align*} 
\end{linenomath} 
$\Phi_{\mathcal{S}}$ is the submatrix of $\Phi$ composed by the columns supported in $\mathcal{S} \subset [m]$. 
\end{propo}
Let the coherence of a matrix $\Phi$ be given by $\eta (\Phi):= \max_{i \neq j} \left| \langle v_i, v_j \rangle  \right|$ defined over distinct pairs of normalized (Euclidean norm) columns of the matrix $\Phi$. The coherence upper bounds the restricted isometry constant, and we use this fact in our proof:
\begin{propo}[Coherence bounds restricted isometry constant]\label{prop:coherence_upper_bounds}
If the matrix $\Phi \in M^{n \times m}$ has $\ell_2$-normalized columns $\{v_1,\dots, v_m\}$,  then
\begin{linenomath}
\begin{align*}
\delta_1 = 0, \quad \delta_2 = \eta, \quad \delta_s \leq \eta (s - 1), s \geq 2.
\end{align*} 
\end{linenomath} 
\end{propo}
\begin{proof}
See proof in \cite{Foucart_mathematical_compressive_sensing}.
\end{proof}

The uniqueness of solutions of the ergodic basis pursuit is a consequence of the following results.

\begin{theorem}[Uniqueness of noiseless recovery \cite{CANDES_2008,Foucart_mathematical_compressive_sensing}]\label{thm:candes_exact_recovery}
Suppose $y = \Phi c$ where $c \in \mathbb{R}^m$ is an $s-$sparse vector. Also, suppose that the $2s$-th restricted isometry constant of the matrix $\Phi \in M^{n \times m}$ satisfies $\delta_{2s} < \sqrt{2} - 1$. Then $c$ is the unique minimizer of
\begin{linenomath} 
\begin{align*}
\min_{u \in \mathbb{R}^m} \|u\|_1 \quad \mathrm{subject~to}~ \Phi u = y.
\end{align*} 
\end{linenomath} 
\end{theorem}
\begin{proof}
See proof in \cite{Candes_2005_decoding,Foucart_mathematical_compressive_sensing}.
\end{proof}
In case of measurement corrupted by noise, the following result holds:
\begin{theorem}[Noisy recovery]\label{thm:appendix_Candes_noise}
Suppose $y = \Phi c + z$ with $\|z\|_{2} \leq \varepsilon$, and denote $c^{\star}$ the solution to the convex minimization problem
\begin{linenomath} \begin{align}\label{eq:Candes_noisy}
    \min_{\tilde{u} \in \mathbb{R}^{m}} \|\tilde{u}\|_1 ~\mathrm{subject~to}~\|y - \Phi \tilde{u}\|_2 \leq \varepsilon.
\end{align}\end{linenomath}
Assume that $\delta_{2s} < \sqrt{2} - 1$. Then the solution to  \eqref{eq:Candes_noisy} obeys
\begin{linenomath}
\begin{align*}
    \|c^{\star} - c\|_2 \leq K_0 s^{-1/2}\|c - c_s\|_1 + K_1 \varepsilon,
\end{align*} 
\end{linenomath} 
for constants $K_0, K_1 > 0$ and $c_s$ denote the vector $c$ with all but the $s-$largest entries set to zero.
\end{theorem}
\begin{proof}
See proof in \cite{CANDES_2008,Foucart_mathematical_compressive_sensing}.
\end{proof}

\subsection{Exponential mixing condition} 
We consider a class of chaotic dynamical systems --- exponentially mixing systems --- that satisfies 
a concentration inequality obtained in \cite{Hang_2017_bernstein-type}. Here, we state this result applied to network dynamics. 

\begin{definition}[Exponential mixing condition]
The network dynamics $(F, \mu)$ satisfies the exponential mixing condition for some constant $\gamma > 0 $ if for all $\psi \in \mathcal{C}^1(M^N;\mathbb{R})$ and $\varphi \in L^1(\mu)$ there exists a constant $K(\psi, \varphi) > 0$ such that
\begin{linenomath} \begin{align}
    \Big|\int_{M^N}\psi \cdot (\varphi \circ F^n) d\mu
- \int_{M^N} \psi d\mu \int_{M^N} \varphi d\mu \Big| \leq K(\psi, \varphi)e^{-\gamma n}, \quad n \geq 0.
\end{align}\end{linenomath}
\end{definition}
We state an adapted version for network dynamics of the concentration inequality \cite{Hang_2017_bernstein-type} for $\mathcal{C}^1(M^N; \mathbb{R})$ observables.

\begin{theorem}[Bernstein inequality for exponential mixing network dynamics \cite{Hang_2017_bernstein-type}.]\label{thm:bernstein_inequality}
Let $(F, \mu)$ be an exponential mixing network dynamical system on $M^N$ for some constant $\gamma > 0$. Moreover, let $\psi \in \mathcal{C}^1(M^N; \mathbb{R})$ be a function such that $\int_{M^N} \psi d\mu = 0$ and assume that there exist $\varsigma > 0$, $\varkappa > 0$ and $\sigma \geq 0$ such that $\|D\psi\|_{\infty} \leq \varsigma$, $\|\psi\|_{\infty} \leq \varkappa$, and $\|\psi^2\|_{\mu}^2 \leq \sigma^2$. Let $\mathcal{N} \subset \mathbb{N}$ be defined as
\begin{linenomath} 
\begin{align*}
    \mathcal{N} := [3, \infty) \bigcap \Big\{ p \in \mathbb{N} ~:~p^2 \geq \frac{808 (3 \varsigma + \varkappa)}{\varkappa} ~\mathrm{and}~ \frac{p}{(\ln{p})^{2}}  \geq 4 \Big\}.
\end{align*} 
\end{linenomath} 
Then, for all $\varepsilon > 0$ and all
\begin{linenomath} 
\begin{align}\label{eq:n_0_bound}
    n \geq n_0 := \max \Big\{e^{\frac{3}{\gamma}}, \min_{\mathcal{N}} p \Big\},
\end{align}
\end{linenomath}
we have
\begin{linenomath} 
\begin{align}
    \mu \Big( x_0 \in M^N ~:~ \Big|\frac{1}{n} \sum_{k = 0}^{n-1} \psi \circ F^k(x_0) \Big| \geq \varepsilon \Big) &\leq 4 e^{-\theta(n, \varepsilon, \sigma, \varkappa)},
\end{align}
\end{linenomath}
where
\begin{linenomath}
\begin{align*}
 \theta(n, \varepsilon, \sigma, \varkappa) & :=  \frac{n \varepsilon^2}{8(\ln{n})^{\frac{2}{\gamma}}(\sigma^2+\varepsilon \varkappa/3)}. 
\end{align*}
\end{linenomath} 
\end{theorem}

\subsection{Semimetric between probability measures}
\label{ssec:metric_prob_measures}
We consider exponentially mixing systems that have near product structure. To be more precise, we introduce a semimetric between probabilities measures suitable to our results. Let $\mathcal{M}(M^N)$ be the set of probability measures on $M^N$. We introduce a probability semimetric \cite{rachev1991probability} between measures on $\mathcal{M}(M^N)$ over a reference finite set of functions $\mathcal{K}$ that is composed by functions on the given network library $\mathcal{L}$. In other words, elements of $\mathcal{K}$ are of the form $\phi_{ij}^{pq} \circ \pi_{\mathcal{J}}$ with $i, j \in \mathcal{J}  \subset [N]$. They are integrated over a lower dimensional space than the ambient space $M^N$, which motivates to define a semimetric out of it, rather than using other metrics on $\mathcal{M}(M^N)$.

\begin{definition}
For any $\mu, \nu \in \mathcal{M}(M^N)$ we define the semimetric over a reference network library $\mathcal{L}$ as
\begin{linenomath} \begin{align}
d_{\mathcal{K}}(\mu, \nu) = \max_{\psi \in \mathcal{K}} \left| \int_{M^N} \psi d\mu - \int_{M^N} \psi d \nu \right|.
\end{align}\end{linenomath}
\end{definition}
$d_{\mathcal{K}}(\mu, \nu)$ is a semimetric and not a metric because: it is symmetric, it satisfies the triangular inequality, and when $\mu = \nu$ implies that $d_{\mathcal{K}}(\mu, \nu) = 0$ but not the converse. Indeed, consider the set $\mathcal{K}$ given by 
\begin{linenomath} 
\begin{align*}
\mathcal{K} = \{\psi_i: M_i \to \mathbb{R} : i \in [N], \int \psi_i dx_i = 0, \psi_i(0) = 0\},
\end{align*} 
\end{linenomath} 
where we assume that $0 \in M_i$ for any $i \in [N]$. Moreover, let $\delta_{0}$ be the Dirac measure at $0$. Consider the following two product measures
\begin{linenomath} 
\begin{align*}
\mu = \mathrm{Leb}^N \qquad \nu = \delta_{0}^N.
\end{align*} 
\end{linenomath} 
It follows that $d_{\mathcal{K}}(\mu, \nu) = 0$ but $\mu \neq \nu$.

In what follows in Section \ref{sec:EBP_has_unique_sol}, it is useful to consider the following finite set $\mathcal{K} = (\mathcal{L} \cdot \mathcal{L})$, where $(\mathcal{L} \cdot \mathcal{L}) = \{(\psi_i \cdot \psi_j): \psi_i, \psi_j \in \mathcal{L}\}$, removing any redundancy.

\subsection{Orthogonal polynomials}

We recall some results for orthonormal polynomials. First, let us state an inequality for orthonormal polynomials in one variable \cite{Szego_1939,Ftorek_2014}. Here we consider a system of orthonormal polynomials $\{\varphi_p(x)\}_{p \geq 0}$ with respect to a measure $\nu$ that is absolutely continuous to Lebesgue, whose density is $\rho$. Since we are in the one variable case, the index $p$ corresponds to the degree to which the coefficient $x^p$ is positive.

\begin{theorem}[One variable Korous inequality \cite{Szego_1939,Ftorek_2014}]\label{thm:Korous_1d}
Let $\{\varphi_p(x)\}_{p \geq 0}$ be a generalized system of orthonormal polynomials with respect to (w.r.t.) the density $\lambda(x)$ and $\{\tilde{\varphi}_p(x)\}_{p \geq 0}$ be a system of orthonormal polynomials w.r.t. the density $\tilde{\lambda}(x)$ such that 
\begin{linenomath}
\begin{align*}
\lambda(x) = \rho(x) \tilde{\lambda}(x),
\end{align*} 
\end{linenomath} 
be two weight (density) functions on the segment $(a, b)$, where $\rho(x) \geq \rho_0 > 0$ and $\rho$ is Lipschitz with constant $\mathrm{Lip}(\rho)$. Then the following estimation
\begin{linenomath} 
\begin{align}\label{eq:Korous_1d}
|\varphi_p(x)| \leq \frac{1}{\rho_0} |\tilde{\varphi}_p(x)| + \frac{K \mathrm{Lip}(\rho)}{\rho_0^{3/2}}(|\tilde{\varphi}_p(x)| + |\tilde{\varphi}_{p - 1}(x)|),
\end{align}
\end{linenomath}
where $\rho_0 = \min_{x \in (a, b)} \rho(x)$, $x \in (a, b)$ and $K = \max \{|a|, |b|\}$.
\end{theorem}

We also recall a result for the product of orthonormal polynomials \cite{dunkl_xu_2014}. 
\begin{propo}[Proposition 2.2.1 in \cite{dunkl_xu_2014}]\label{propo:pairwise_orthogonal_polynomials}
Let $\rho(x_1, x_2) = \rho_1(x_1)\rho_2(x_2)$, where $\rho_1$ and $\rho_2$ are two weight functions of one variable. Let $\{\varphi_{1}^p(x_1)\}_{p \geq 0}^{\infty}$ and $\{\varphi_2^{q}(x_2)\}_{q \geq 0}^{\infty}$ with $p,q \in \mathbb{N}$ be sequences of orthogonal polynomials with respect to $\rho_1$ and $\rho_2$, respectively. Then a mutually orthogonal basis of the space of orthogonal polynomials of degree $r$ with respect to $\rho$ is given by:
\begin{linenomath} 
\begin{align*}
\varphi_{12}^{pq}(x_1, x_2) = \varphi_{1}^p(x_1)\varphi_2^{q}(x_2), \quad 0 \leq p + q \leq r.
\end{align*} 
\end{linenomath} 
Furthermore, if $\{\varphi_{1}^p(x_1)\}_{p \geq 0}^{\infty}$ and $\{\varphi_2^{q}(x_2)\}_{q \geq 0}^{\infty}$ are orthonormal with respect to $\rho_1$ and $\rho_2$, respectively, then so is $\varphi_{12}^{pq}(x_1, x_2)$ with respect to $\rho$.
\end{propo}

\subsection{Gram-Schmidt process}

Let $\nu$ be a measure on $M^N$ that is absolutely continuous with respect to Lebesgue. We address the problem of ortho-normalizing the ordered network library $\mathcal{L}$ with respect to a measure $\nu$. Let us denote the inner product w.r.t. $\nu$ as
\begin{linenomath} \begin{align}\label{eq:inner_product}
        \langle \phi_k, \phi_l  \rangle = \int_{M^{N}} \phi_k \phi_l d \nu \qquad \|\phi_l\|_{\nu}^2 = \langle \phi_l, \phi_l \rangle.
    \end{align}\end{linenomath}
We consider the Gram-Schmidt (GS) process, which is a recursive method given as
\begin{linenomath} \begin{align}\label{eq:GS_process}
\begin{split}
\hat{\varphi}_1 &= \phi_1 \\
\hat{\varphi}_{k + 1} &= \phi_{k + 1} - \sum_{l = 1}^{k} \langle\phi_{k + 1}, \varphi_l \rangle \varphi_l, \\ 
\varphi_k &:= \frac{\hat{\varphi}_k}{\|\hat{\varphi}_k\|_{\nu}},
 \quad k \geq 1.
\end{split}
\end{align}\end{linenomath}
From the ordered network library $\mathcal{L}$ the induced library $\mathcal{L}_{\nu} = \{\varphi_k: M^N \to \mathbb{R}: k \in [m]\}$ is given by each $k$-th orthonormal function written as a linear combination, whose coefficients are projections on the preceding orthonormal functions.

\section{Network library is preserved under Gram-Schmidt process}

To ensure that the ergodic basis pursuit has a unique solution, the library matrix used in the reconstruction must satisfy the restricted isometry property, as defined in Equation \eqref{eq:RIP_property}. However, a priori, the library matrix associated with the network library $\mathcal{L}$, in which $F_{\alpha}$ has a sparse representation, does not satisfy RIP. Our strategy is to introduce a new library $\mathcal{L}_{\nu}$ that is orthonormal with respect to a suitable measure $\nu$ in $L^2(\nu)$, where the associated library $\Phi_{\nu}(X)$ satisfies RIP. 

\subsection{The set of pairwise polynomials of degree at most $r$}

We consider a network library given by polynomials in $N$ variables of degree at most $r$. This also can be applied to trigonometric polynomials in $N$ variables. 

Given $r \geq 2$, let us denote the exponent vector set 
\begin{linenomath} \begin{align}\label{eq:exp_vec_set}
\mathcal{V}_r := \{(p, q) \in [r - 1]^2:p + q \leq r\},
\end{align}\end{linenomath}
which is organized in graded lexicographic order and denoted as $(p^{\prime}, q^{\prime}) \prec (p, q)$. Moreover, denote
\begin{linenomath}
\begin{align*}
\mathcal{I}_{r} &= \{\phi_{i}^{p}(x_i) = x_i^p~:~i \in [N], p \in [r]\}, \\
\mathcal{P}_r &= \{\phi_{ij}^{pq}(x_i, x_j) = x_i^p x_j^q~:~ i, j \in [N], i \neq j, (p, q) \in \mathcal{V}_r\},
\end{align*} 
\end{linenomath} 
where we remove any redundancy. We can unify the notation for both if we denote elements of $\mathcal{I}_{r}$ as $\phi_{i0}^{p0}(x_i, x_j) = x_i^p$. We define the set of pairwise polynomials in $N$ variables with a degree at most $r$
\begin{linenomath} 
\begin{align*}%\label{eq:homogeneous_polynomials}
\begin{split}
\mathcal{L} &= \{1\} \cup \mathcal{I}_r \cup \mathcal{P}_r \\
&= \{\phi_{ij}^{pq}(x_i, x_j) = x_i^p x_j^q ~:~i \in [N],j \in \{0\}\cup[N], i \neq j\\ 
& \hspace{4.25cm} p = \{0\}\cup[r], q \in \{0\} \cup[r - 1],  \\ 
&\hspace{4.25cm} p + q \leq r\} ,
\end{split}
\end{align*} 
\end{linenomath}  
whose cardinality is given by $m = \binom{N}{2} \binom{r}{2} + N r + 1$. In fact, the independent polynomial $1$ contributes with one term. The cardinality of $\mathcal{I}_r$ is $Nr$ because for each $i \in [N]$ there are $r$ polynomials in the subset $\{\varphi_{i0}^{p0}\}_{p \in [r]}$. Finally, for $\mathcal{P}_r$ fix a pair $i, j \in [N]$ with $i \neq j$. For each pair, the degree of the pairwise polynomial is $p + q = d \in [r]$. Since they are constrained through their sum, for each degree $d \in [r]$, the first component in the sum $p \in \{1, \dots, d - 1\}$, which also determines the value of $q$ correspondingly. Then, there are total of $\sum_{d = 1}^r d - r$ possible combinations. Rewriting it
\begin{linenomath} 
\begin{align*}
\sum_{d = 1}^r d - r &= \frac{r(r + 1)}{2} - r\\
 &= \frac{r(r - 1)}{2} \\
&= \binom{r}{2}.
\end{align*} 
\end{linenomath}   
Running over all possible distinct pairs $i, j$, we obtain the total cardinality of $\mathcal{P}_r$ equal to $\binom{N}{2} \binom{r}{2}$.

Here we adopt the following ordering: fix $j, q = 0$ and start with $p = 0$. Then, for each $i \in [N]$, we run through $p \in [r]$, covering all monomials that depend on one variable. Subsequently, for each element in $\{(i, j) \in [N]^2: i \in [N], j = i+1, \dots, N\}$ (organized in lexicographic order), we run through the exponent vector set $\mathcal{V}_r$.

\subsection{Network library is preserved}
\label{sec:net_is_preserved}

Given a trajectory $\{x(t)\}_{t = 0}^n$ that is sampled from $\mu_{\alpha}$, the natural choice would be to orthonormalize with respect to $\mu_{\alpha}$ itself. However, it does not necessarily preserve the sparsity of the representation of $F_{\alpha}$ in the network library $\mathcal{L}$. The next theorem states that the GS process over $\mathcal{L}$ with respect to a product measure $\nu$ introduces a new network library $\mathcal{L}_{\nu}$, and also, $F_{\alpha}$ is still sparsely represented in $\mathcal{L}_{\nu}$. In this new basis, the sparsity level depends on the maximum degree $r$ and the sparsity level of the representation in $\mathcal{L}$. 

Denote the product measure as $\nu = \prod_{i = 1}^N \nu_i$ and denote
\begin{linenomath} \begin{align}\label{eq:prod_moments}
\mathbb{E}(x_i^{p}) &= \int_{M} x_i^{p} d \nu_i(x_i), \quad i \in [N].
\end{align}\end{linenomath}
Consider the following 
\begin{theorem}[Network library is preserved]\label{thm:net_lib_preserved} 
Let $\nu$ be a product measure on $M^N$ that is absolutely continuous with respect to the Lebesgue measure. Gram-Schmidt process maps an $s$-sparse representation of $F_{\alpha}$ in the network library $\mathcal{L}$ to an $\omega_r(s)$-sparse representation in the orthonormal network library $\mathcal{L}_{\nu}$ in $L^2(\nu)$, with 
\begin{linenomath} \begin{align}\label{eq:omega_r_s}
\omega_r(s) = \Big( \floor*{\frac{r}{2}} \big(r - \floor*{\frac{r}{2}} \big)  + r + 1\Big) s.
\end{align}\end{linenomath}
\end{theorem}

We divide the proof into two parts: first, we show that the GS process maps the network library $\mathcal{L}$ to another network library $\mathcal{L}_{\nu}$ that is orthonormal w.r.t. $\nu$. The second part is to calculate the sparsity level of the representation of $F_{\alpha}$ in $\mathcal{L}_{\nu}$.

\subsubsection{Proof of Theorem \ref{thm:net_lib_preserved}}
When we perform the GS process in $L^2(\nu)$ as in  \eqref{eq:GS_process}, to orthonormalize $\mathcal{L}$ with respect to the measure $\nu$, the first element in $\mathcal{L}_{\nu}$ is evidently $1$. Following the order in the network library $\mathcal{L}$ in \eqref{eq:homogeneous_polynomials}, we can show that a general form of all polynomials that depend on only one variable is given by the proposition below.

\begin{propo}[Formula of orthonormal functions in one variable]\label{propo:one_variable_functions}
Let $i \in [N]$ and $p \in [r]$. Then any $\varphi_{i0}^{p0} \in \mathcal{L}_{\nu}$ is given by
\begin{linenomath} \begin{align}\label{eq:general_isolated_function}
\begin{split}
\hat{\varphi}_{i0}^{p0}(x_i) &= x_i^p - \mathbb{E}(x_i^p) - \sum_{l = 1}^{p - 1} \langle x_i^p, \varphi_{i0}^{l0} \rangle \varphi_{i0}^{l0}(x_i), \\
\varphi_{i0}^{p0}(x_i) &= \frac{\hat{\varphi}_{i0}^{p0}(x_i)}{\|\hat{\varphi}_{i0}^{p0}\|_{\nu}},
\end{split}
\end{align}\end{linenomath}
and 
\begin{linenomath} \begin{align}\label{eq:orthogonal_moments_vanish}
\mathbb{E}(\hat{\varphi}_{i0}^{p0}(x_i)) = \mathbb{E}(\varphi_{i0}^{p0}(x_i)) = 0.
\end{align}\end{linenomath}
\end{propo} 
We prove this statement in two parts. First, we continue the Gram-Schmidt process over the ordering of $\mathcal{L}$. So, fix $i = 1$ and run over $p \in [r]$. The next term after the constant function $1$ is 
\begin{linenomath} 
\begin{align*}
\hat{\varphi}_{10}^{10}(x) &= \phi_{10}^{10}(x_1) - \langle \phi_{10}^{10}, 1 \rangle 1 \\
&= x_1 - \mathbb{E}(x_1),
\end{align*} 
\end{linenomath} 
and consequently, $\varphi_{10}^{10}(x_1) = \frac{\hat{\varphi}_{10}^{10}(x_1)}{\|\hat{\varphi}_{10}^{10}\|_{\nu}}$, which satisfies \eqref{eq:general_isolated_function} and \eqref{eq:orthogonal_moments_vanish}. To calculate the next element $\varphi_{10}^{20}(x_1)$, we follow \eqref{eq:GS_process}:
\begin{linenomath} 
\begin{align*}
\hat{\varphi}_{10}^{20}(x_1) &= \phi_{10}^{20}(x_1) - \langle \phi_{10}^{20}, 1 \rangle 1 - \langle \phi_{10}^{20}, \varphi_{10}^{10} \rangle \varphi_{10}^{10}(x_1) \\
&= x_1^2 - \mathbb{E}(x_1^{2}) - \langle x_1^2, \varphi_{10}^{10} \rangle \varphi_{10}^{10} (x_1),
\end{align*} 
\end{linenomath} 
and consequently, $\varphi_{10}^{20}(x_1) = \frac{\hat{\varphi}_{10}^{20}(x_1)}{\|\hat{\varphi}_{10}^{20}\|_{\nu}}$. Following the ordering, we run over all functions of the form $\varphi_{10}^{p0}$, repeating the Gram-Schmidt process \eqref{eq:GS_process} to show that they satisfy \eqref{eq:general_isolated_function} and  \eqref{eq:orthogonal_moments_vanish}.

The next functions involve coordinates that are different from $i = 1$. To prove that these functions satisfy \eqref{eq:general_isolated_function} and \eqref{eq:orthogonal_moments_vanish}, we run a recursive argument. Fix $i = 1$ and $j = 2$, and let us consider the orthogonal function for $p \in [r]$ using Gram-Schmidt process:
\begin{linenomath} 
 \begin{align*}
\hat{\varphi}_{20}^{p0}(x_2) = x_2^p - \mathbb{E}(x_2^p) - \sum_{l = 1}^{r} \langle x_2^p, \varphi_{10}^{l0} \rangle \varphi_{10}^{l0}(x_1) - \sum_{l = 1}^{p - 1} \langle x_2^p, \varphi_{20}^{l0} \rangle \varphi_{20}^{l0}(x_2).
\end{align*}
 \end{linenomath} 
Note that if all inner products of the form $ \langle x_2^p, \varphi_{10}^{l0} \rangle$ are zero, above equation satisfies \eqref{eq:general_isolated_function}. We state the following lemma:

\begin{lemma}\label{lemma:general_inner_prod_1_var}
Let $i, j \in [N]$, $p, q \in [r]$. Suppose that $\varphi_{i0}^{p0}$ is an orthonormal polynomial with respect to $\nu$, i.e., it satisfies \eqref{eq:general_isolated_function} and $\varphi_{i0}^{p0} \in \mathcal{L}_{\nu}$. Then,
\begin{linenomath}
\begin{align*}
\langle \phi_{j0}^{q0}, \varphi_{i0}^{p0} \rangle = 0.
\end{align*} 
\end{linenomath}  
whenever $i \neq j$.
\end{lemma}
\begin{proof}
By Fubini's theorem, we have that
\begin{linenomath} 
\begin{align*}
    \langle \phi_{j0}^{q0}, \varphi_{i0}^{p0} \rangle &=  \int_{M^{N}} \phi_{j0}^{q0}(x_{j})\varphi_{i0}^{p0}(x_{i}) d \nu(x_1, \dots, x_N) \\
     &= \langle \phi_{j0}^{q0}, 1 \rangle \langle  \varphi_{i0}^{p0}, 1 \rangle.
\end{align*} 
\end{linenomath} 
Since $\varphi_{i0}^{p0}$ satisfies \eqref{eq:general_isolated_function}, it is orthonormal to $1$, and the claim holds.
\end{proof}
We use above Lemma \ref{lemma:general_inner_prod_1_var} to the inner product $ \langle x_2^p, \varphi_{10}^{l0} \rangle$, where $\varphi_{10}^{l0}$ satisfies  \eqref{eq:general_isolated_function}. We conclude that for any $p \in [r]$: $\varphi_{20}^{p0}$ also satisfies \eqref{eq:general_isolated_function} and \eqref{eq:orthogonal_moments_vanish}. We run iteratively, choosing $i \geq 2$ and $j = i + 1$, and repeating the argument to conclude the proof of Proposition \ref{propo:one_variable_functions}.

For polynomials involving two variables, it is enough to construct them from the orthonormal polynomials in one variable as follows:
\begin{propo}[Formula of orthonormal functions in two variables]\label{propo:general_pairwise_function}
Let $r \geq 2$, $i, j \in [N]$ with $i \neq j$ and $(p, q) \in \mathcal{V}_r$. Then 
\begin{linenomath} \begin{align}\label{eq:general_pairwise_function}
\varphi_{ij}^{pq}(x_i, x_j) = \varphi_{i0}^{p0}(x_i)\varphi_{j0}^{q0}(x_j).
\end{align}\end{linenomath}
\end{propo}
\begin{proof}
The measure $\nu = \prod_{i = 1}^N \nu_i$. For each marginal $\nu_i$, let $\rho_i$ be the density function. Then, we apply Proposition \ref{propo:pairwise_orthogonal_polynomials} for every distinct pair of nodes $i, j \in [N]$. 
\end{proof}
To construct $\mathcal{L}_{\nu}$ we combine Proposition \ref{propo:one_variable_functions} and Proposition \ref{propo:general_pairwise_function}. The Gram-Schmidt process induces a set of orthonormal polynomials in one variable that satisfies the ordering of $\mathcal{L}$. The ordering of polynomials in two variables in $\mathcal{L}_{\nu}$ also satisfies, by construction, the ordering in $\mathcal{L}$. This proves the first part of Theorem \ref{thm:net_lib_preserved}.

To prove the second part of the theorem, we also use that $\mathcal{L}_{\nu}$ is constructed via the Gram-Schmidt process. Let $u, u_{\nu} \in \mathbb{R}^{m}$ be vectors  with $m = \binom{N}{2} \binom{r}{2} + N r + 1$ given by
\begin{linenomath} 
\begin{align*}
u & = (1, x_1, \dots, x_1^r, x_2, \dots, x_2^r, \dots, x_1 x_2, \dots,  x_{N - 1}x_N^r)
\end{align*} 
\end{linenomath} 
and
\begin{linenomath} 
\begin{align*}
u_{\nu} &= (1, \varphi_{10}^{10}(x_1), \dots, \varphi_{10}^{r0}(x_1), \varphi_{20}^{10}(x_2), \dots, \varphi_{20}^{r0}(x_2), \dots , \\ 
&\qquad \qquad \varphi_{12}^{11}(x_1, x_2), \dots, \varphi_{N-1, N}^{r - 1, 1}(x_{N - 1}, x_N)).
\end{align*} 
\end{linenomath} 
Each coordinate of $u$ is an element of $\mathcal{L}$ that can be written as a linear combination of elements in $\mathcal{L}_{\nu}$. In fact, we rewrite \eqref{eq:general_isolated_function} as
\begin{linenomath} 
\begin{align}\label{eq:x_i_p_in_basis}
x_i^p = \|\hat{\varphi}_{i0}^{p0}\|_{\nu} \varphi_{i0}^{p0}(x_i) + \mathbb{E}(x_i^p) + \sum_{l = 1}^{p - 1} \langle x_i^p, \varphi_{i0}^{l0} \rangle \varphi_{i0}^{l0}(x_i),
\end{align}
\end{linenomath} 
which expresses the polynomials in one variable as a linear combination of orthonormal polynomials in one variable. For the two variables polynomials of the form $x_i^p x_j^q$, we replace each term in the multiplication by \eqref{eq:x_i_p_in_basis} and
\begin{enumerate}
\item Replace any multiplication of orthonormal polynomial of the form $\varphi_{i0}^{p0}(x_i) \varphi_{j0}^{q0}(x_j)$ by the orthonormal polynomial in two variables Equation \eqref{eq:general_pairwise_function}. 
\item Use these identities that follow from Fubini's theorem:
 \begin{linenomath}
 \begin{align*}
\langle x_i^p, \varphi_{i0}^{l0} \rangle  \langle x_j^q, \varphi_{j0}^{k0} \rangle =  \langle x_i^p x_j^q, \varphi_{ij}^{lk} \rangle, 
\end{align*} \end{linenomath} 
\begin{linenomath} \begin{align*}
\mathbb{E}(x_i^p) \langle x_j^q, \varphi_{j0}^{k0} \rangle = \langle x_i^p x_j^q, \varphi_{j0}^{k0} \rangle
\end{align*} \end{linenomath} 
and 
\begin{linenomath} \begin{align*}
\|\hat{\varphi}_{i0}^{p0}\|_{\nu}\|\hat{\varphi}_{j0}^{q0}\|_{\nu} = \|\hat{\varphi}_{ij}^{pq}\|_{\nu}.
\end{align*} 
\end{linenomath} 
\end{enumerate}
Then, we can recast the Gram-Schmidt process as the following linear equation
\begin{linenomath} \begin{align}\label{eq:linear_equation_functions}
u^T = u_{\nu}^T \mathbf{R}_{\nu},
\end{align}\end{linenomath}
where $T$ denotes the transpose and $\mathbf{R}_{\nu} \in \mathbb{R}^{m \times m}$ is a triangular matrix given as
\begin{linenomath} \begin{align}\label{eq:R_nu}
\mathbf{R}_{\nu} = \begin{pmatrix}
1 & \mathbf{V}_1 & \mathbf{V}_2  \\
0 & \mathbf{U}_1 & \mathbf{U}_{2}^1 \\
0 & 0 & \mathbf{U}_{2}
\end{pmatrix}.
\end{align}\end{linenomath}
Here $\mathbf{V}_1  \in \mathbb{R}^{rN} $ and $\mathbf{V}_2 \in \mathbb{R}^{\binom{N}{2} \binom{r}{2}}$ are given by
\begin{linenomath} \begin{align*}
\mathbf{V}_1 = \begin{pmatrix}
v_1 & v_2 & \dots & v_N
\end{pmatrix} \qquad 
\mathbf{V}_2 = \begin{pmatrix}
v_{12} & v_{13} & \dots & v_{N-1, N}
\end{pmatrix},
\end{align*} \end{linenomath} 
where for each $i, j \in [N]$ with $i \neq j$, $v_i  \in \mathbb{R}^{r} $ and $v_{ij} \in \mathbb{R}^{\binom{r}{2}}$:
\begin{linenomath} \begin{align*}
v_i = \Big(
\mathbb{E}(x_i), \dots, \mathbb{E}(x_i^r)
\Big)
\end{align*} \end{linenomath} 
and
\begin{linenomath} \begin{align*}
v_{ij} = \Big(
\mathbb{E}(x_i) \mathbb{E}(x_j), \mathbb{E}(x_i) \mathbb{E}(x_j^2), \mathbb{E}(x_i^2) \mathbb{E}(x_j), \dots, \mathbb{E}(x_i^{r - 1})\mathbb{E}(x_j) \Big).
\end{align*} \end{linenomath} 
Also, $\mathbf{U}_1 \in \mathbb{R}^{Nr \times Nr}$ and $\mathbf{U}_2 \in \mathbb{R}^{\binom{N}{2}  \binom{r}{2} \times \binom{N}{2}  \binom{r}{2}}$  are block diagonal matrices defined as follows:
\begin{linenomath} \begin{align*}
\mathbf{U}_1 = \mathrm{diag}\Big(U_{1}, \dots, U_{N}\Big) \qquad
\mathbf{U}_2 = \mathrm{diag}\Big(U_{12}, \dots, U_{N - 1, N}\Big),
\end{align*} \end{linenomath} 
where for each $i, j \in [N]$, $U_{i} \in \mathbb{R}^{r \times r}$ and $U_{ij} \in \mathbb{R}^{\binom{r}{2} \times \binom{r}{2}}$ are given by
\begin{linenomath} \begin{align*}
U_{i} &= \begin{pmatrix}
\|\hat{\varphi}_{i0}^{10}\|_{\nu} & \langle x_i^2, \varphi_{i0}^{10} \rangle & \dots & \langle x_i^r, \varphi_{i0}^{10} \rangle \\
0 & \|\hat{\varphi}_{i0}^{20}\|_{\nu} & \dots & \langle x_i^r, \varphi_{i0}^{20}\rangle \\
\vdots & \ddots & \ddots & \vdots & \\
0 & \dots  &0 &  \|\hat{\varphi}_{i0}^{r0}\|_{\nu} & 
\end{pmatrix}  \\
U_{ij} &= \begin{pmatrix}
\|\hat{\varphi}_{ij}^{11}\|_{\nu} &  \|\hat{\varphi}_{i0}^{10}\|_{\nu} \langle x_i, \varphi_{j0}^{10} \rangle &  \|\hat{\varphi}_{j0}^{10}\|_{\nu} \langle x_i^2, \varphi_{i0}^{10} \rangle & \dots & \|\hat{\varphi}_{j0}^{10}\|_{\nu} \langle x_i^{r - 1}, \varphi_{i0}^{10} \rangle \\
0 & \|\hat{\varphi}_{ij}^{12}\|_{\nu} & 0 &  \dots & 0 \\
0 & 0 & \|\hat{\varphi}_{ij}^{21}\|_{\nu} &  \dots & \|\hat{\varphi}_{j0}^{10}\|_{\nu} \langle x_i^{r - 1}, \varphi_{i0}^{20} \rangle \\
\vdots & \vdots & 0 & \ddots & \vdots & \\
0 & \dots  &0 &0 &  \|\hat{\varphi}_{ij}^{r - 1, 1}\|_{\nu} & 
\end{pmatrix},
\end{align*} \end{linenomath} 
and $\mathbf{U}_2^1 \in \mathbb{R}^{N r \times \binom{N}{2}  \binom{r}{2}}$ is a block matrix
\begin{linenomath} \begin{align*}
\mathbf{U}_2^1 = 
\begin{pmatrix}
U_{12}^{1} & U_{13}^1 & \dots & 0 \\
U_{12}^{2} & 0 & \dots & 0\\
0 & U_{13}^3 & \dots & \vdots \\
\vdots & \ddots & \ddots & U_{N - 1, N}^{N - 1}\\
0 & \dots  &0 &  U_{N - 1,N}^N  
\end{pmatrix}, 
\end{align*} \end{linenomath} 
where for each $i \in [N]$, $j = i, \dots, N$:
\begin{linenomath} \begin{align*}
U_{ij}^i = \begin{pmatrix}
\|\hat{\varphi}_{i0}^{10}\|_{\nu} \mathbb{E}(x_j)  &  \|\hat{\varphi}_{i0}^{10}\|_{\nu} \mathbb{E}(x_j^2) &  \langle x_i^2 x_j, \varphi_{i0}^{10} \rangle & \dots & \langle x_i^{r - 1} x_j, \varphi_{i0}^{10} \rangle \\
0 & 0 & \|\hat{\varphi}_{i0}^{20}\|_{\nu} \mathbb{E}(x_j) & \dots & \langle x_i^{r - 1} x_j, \varphi_{i0}^{20} \rangle \\
\vdots & \vdots & & \ddots & \vdots \\
0 & 0 & & \dots & \|\hat{\varphi}_{i0}^{(r-1)0}\|_{\nu} \mathbb{E}(x_j) 
\end{pmatrix} \in \mathbb{R}^{r \times \binom{r}{2}} \\
U_{ij}^j = \begin{pmatrix}
 \mathbb{E}(x_i)\|\hat{\varphi}_{j0}^{10}\|_{\nu}  & \langle x_i x_j^2, \varphi_{j0}^{10} \rangle  & \mathbb{E}(x_i^2) \|\hat{\varphi}_{j0}^{10}\|_{\nu}  & \dots &  \mathbb{E}(x_i^{r - 1}) \|\hat{\varphi}_{j0}^{10}\|_{\nu}  \\
0 & \|\hat{\varphi}_{i0}^{10}\|_{\nu} \mathbb{E}(x_j^2) & 0 & \dots & 0 \\
\vdots & \vdots &  \vdots & \ddots & \vdots \\
0 & 0 & 0&\dots & 0
\end{pmatrix} \in \mathbb{R}^{r \times \binom{r}{2}}.
\end{align*} \end{linenomath} 
Linear equation \eqref{eq:linear_equation_functions} is valid for every point $x \in M^N$. Hence, evaluating along the trajectory $\{x(t)\}_{t = 0}^n$ we obtain:
\begin{linenomath} \begin{align}\label{eq:change_of_set_of_basis}
\Phi(X) = \Phi_{\nu}(X) \mathbf{R}_{\nu}. 
\end{align}\end{linenomath}
Consider an $s$-sparse representation in $\mathcal{L}$, then there is an $s$-sparse vector $c \in \mathbb{R}^{m}$ such that
\begin{linenomath} \begin{align*}
\bar{x} = \Phi(X) c,
 \end{align*} \end{linenomath} 
where we dropped the dependence on the node $i \in [N]$ for a moment. Note that $\mathcal{L}_{\nu}$ is also a set of basis functions, so there exists a $c_{\nu}$ such that
\begin{linenomath} \begin{align*}
\Phi(X) c = \Phi_{\nu}(X)c_{\nu}. 
\end{align*} \end{linenomath} 
 \eqref{eq:change_of_set_of_basis} implies that $c_{\nu} = \mathbf{R}_{\nu} c$. Since $c_{\nu}$ is the linear combination of $s$ columns of $\mathbf{R}_{\nu}$, the sparsity level of $c_{\nu}$ is given by the number of non-zero entries of $\mathbf{R}_{\nu}$ multiplied by the sparsity level $s$ of $c$. 

The sparsity of $\mathbf{R}_{\nu}$ columns can be upper bounded by counting the non-zero entries of columns in the block matrices involving the pairwise interaction. It is enough to calculate the maximum number of elements in the multiplication $x_i^{p} x_j^{q}$ using \eqref{eq:x_i_p_in_basis} for all combinations of $p, q \in [r - 1]$ with $p + q \leq r$. More precisely, for a $p$ in \eqref{eq:x_i_p_in_basis} there is a linear combination of $p + 1$ elements of $\mathcal{L}_{\nu}$. Then, in the multiplication $x_i^{p} x_j^{q}$ there are at maximum
\begin{linenomath} \begin{align*}
\omega_r = \max_{p, q \in [r - 1], p + q \leq r}(p + 1)(q + 1),
\end{align*} \end{linenomath} 
which has the following expression
\begin{linenomath} \begin{align*}
\omega_r = \floor*{\frac{r}{2}} \big(r - \floor*{\frac{r}{2}} \big)  + r + 1.
\end{align*} \end{linenomath} 
So, $c_{\nu}$ is an $\omega_r(s)$-sparse vector with $\omega_r(s) = \Big( \floor*{\frac{r}{2}} \big(r - \floor*{\frac{r}{2}} \big)  + r + 1\Big) s$.

We repeat the same argument for each $i \in [N]$ separately, concluding the proof of Theorem \ref{thm:net_lib_preserved}.

\subsection{Bounds for orthonormal polynomials}
\label{sec:bounds_ortho}
In the next section, we will need bounds of orthonormal polynomials with respect to the product measure $\nu$. We focus on the one variable case because, as we have seen in the previous section, it suffices to analyze this case.

First, note that: consider a system of orthonormal polynomials $\{\psi_p(z)\}_{p \geq 0}$ with weight (density) function $\lambda(z)$ defined on the interval $[a_2, b_2] \subset \mathbb{R}$. The linear transformation $T(x) = \alpha x + \beta$ with $\alpha \neq 0$ maps an interval $[a_1, b_1]\subset \mathbb{R}$ onto the interval $[a_2, b_2]$, and $\lambda \circ T(x)$ into $\lambda$, then the polynomials 
\begin{linenomath} \begin{align*}
\{\mathrm{sgn}(\alpha)^p |\alpha|^{\frac{1}{2}} \psi_p \circ T(x)\}_{p \geq 0}
\end{align*} \end{linenomath} 
are orthonormal on $[a_1, b_1]$ with the weight function $\lambda \circ T(x)$.

Consider the set of Legendre polynomials $\{L_p(z)\}_{p \geq 0}$ which is defined on $[-1, 1]$ with $\lambda(z) = 1$. From the above remark, any Legendre polynomial $\{\hat{L}_p(x)\}_{p \geq 0}$ defined in an arbitrary interval $[a, b]$ is given by
\begin{linenomath} \begin{align}\label{eq:legendre_extended_poly}
\Big\{\hat{L}_p(x) := \mathrm{sgn}(\frac{2}{b - a})^p \left|\frac{2}{b - a}\right|^{\frac{1}{2}} L_p\big(\frac{2}{b - a}(x - b) + 1\big)\Big\}_{p \geq 0}
\end{align}\end{linenomath}
with weight $\lambda\big(\frac{2}{b - a}(x - b) + 1\big) = 1$.
Note that $\|L_p\|_{\infty} \leq 1$ \cite{Szego_1939}, consequently, 
\begin{linenomath} \begin{align*}
\|\hat{L}_p\|_{\infty} \leq \Big(\frac{2}{b - a}\Big)^{\frac{1}{2}} \|L_p\|_{\infty} \leq \Big(\frac{2}{b - a}\Big)^{\frac{1}{2}}.
\end{align*} \end{linenomath} 
We apply the above observation to our case, using the Korous inequality for orthonormal polynomials \ref{thm:Korous_1d}. See the following:
\begin{propo}[Supremum norm of orthonormal polynomials in one variable]\label{propo:one_variable_sup}
For a given $i \in [N]$ let $M_i = [a, b] \subset \mathbb{R}$ with $b > a$. Consider the one variable orthonormal polynomials $\{\varphi_{i0}^{p0}(x_i)\}_{p \in [r]}$ with respect to $\nu_i$, which is the one-dimensional marginal of the product measure $\nu$. Suppose that $\nu_i$ is absolutely continuous with respect to Lebesgue and its density $\rho_i$ is at least Lipschitz with constant $\mathrm{Lip}(\rho_i)$. Moreover, $\rho_i(x_i) > 0$ for any $x_i \in M_i$. The following holds:
\begin{linenomath} \begin{align*}
\|\varphi_{i0}^{p0}\|_{\infty} \leq \Big( \frac{1}{\rho_{0}} + 2\frac{a_1 \mathrm{Lip}(\rho)}{\rho_{0}^{3/2}}\Big) \Big(\frac{2}{b - a}\Big)^{\frac{1}{2}}, \quad p \in [r],
\end{align*} \end{linenomath} 
where $\rho_{0} = \min_{i \in [N]} \{\min_{x \in [a, b]} \rho_i(x)\}$, $a_1 = \max\{|a|, |b|\}$ and $\mathrm{Lip}(\rho) = \max_{i \in [N]}\mathrm{Lip}(\rho_i)$. Moreover,
\begin{linenomath} \begin{align*}
\|D \varphi_{i0}^{p0}\|_{\infty} &\leq \Big( \frac{1}{\rho_{0}} + 2\frac{a_1 \mathrm{Lip}(\rho)}{\rho_{0}^{3/2}}\Big) \Big(\frac{2}{b - a}\Big) r^2.
\end{align*} \end{linenomath} 
\end{propo}
\begin{proof}
Consider the system $\{\hat{L}_p(x)\}_{p \in [r]}$ of Legendre polynomials as in \eqref{eq:legendre_extended_poly} defined on $M_i$. Also, consider the orthonormal polynomials $\{\varphi_{i0}^{p0}(x_i)\}_{p \in [r]}$ with respect to $\nu_i$, which is given by $d\nu_i(x) = \rho_i(x_i) d\mathrm{Leb}(x_i) = \rho_i(x_i) \lambda\big(\frac{2}{b - a}(x_i - b) + 1\big) dx_i $. Then, we apply Korous inequality for orthonormal polynomials \ref{thm:Korous_1d}. Additionally, using Markov’s inequality for polynomials
\begin{linenomath} \begin{align*}
\|D \varphi_{i0}^{p0}\|_{\infty} &\leq \Big(\frac{2}{b - a}\Big) r^2 \|\varphi_{i0}^{p0}\|_{\infty},
\end{align*} \end{linenomath} 
and the result holds.
\end{proof}
\begin{coro}[Supremum norm of orthonormal polynomials in two variables]\label{coro:two_variables_sup}
Let $r \geq 2$, $i, j \in [N]$ with $i \neq j$ and $(p, q) \in \mathcal{V}_r$. Then 
\begin{linenomath} \begin{align*}
\|\varphi_{ij}^{pq}\|_{\infty} \leq \Big( \frac{1}{\rho_{0}} + 2\frac{a_1 \mathrm{Lip}(\rho)}{\rho_{0}^{3/2}}\Big)^2 \Big(\frac{2}{b - a}\Big)
\end{align*} \end{linenomath} 
and
\begin{linenomath} \begin{align*}
\|D\varphi_{ij}^{pq}\|_{\infty} \leq 2 \Big( \frac{1}{\rho_{0}} + 2\frac{a_1 \mathrm{Lip}(\rho)}{\rho_{0}^{3/2}}\Big)^2 \Big(\frac{2}{b - a}\Big)^{\frac{3}{2}}r^2.
\end{align*} \end{linenomath} 
\end{coro}
\begin{proof}
\begin{linenomath} \begin{align*}
\|\varphi_{ij}^{pq}\|_{\infty} = \sup_{x_i, x_j \in (a, b)} |\varphi_{ij}^{pq}(x_i, x_j)| \leq  \|\varphi_{i0}^{p0}\|_{\infty}\|\varphi_{j0}^{q0}\|_{\infty},
\end{align*} \end{linenomath} 
and for the derivative, we calculate
\begin{linenomath} \begin{align*}
\|D\varphi_{ij}^{pq}\|_{\infty} = \sup_{x_i, x_j \in (a, b)} |D\varphi_{ij}^{pq}(x_i, x_j)| \leq  \|D\varphi_{i0}^{p0}\|_{\infty}\|\varphi_{j0}^{q0}\|_{\infty} + \|\varphi_{i0}^{p0}\|_{\infty}\|D\varphi_{j0}^{q0}\|_{\infty}.
\end{align*} \end{linenomath} 
The result holds applying Proposition \ref{propo:one_variable_sup}. 
\end{proof}
From here on, for short notation, we denote 
\begin{linenomath} \begin{align}\label{eq:constant_K_rho}
K = K(\mathrm{Lip}(\rho), \rho_0) \equiv \Big( \frac{1}{\rho_{0}} + 2\frac{a_1 \mathrm{Lip}(\rho)}{\rho_{0}^{3/2}}\Big)^2 \Big(\frac{2}{b - a}\Big).
\end{align}\end{linenomath}

\section{Ergodic Basis Pursuit has a unique solution}
\label{sec:EBP_has_unique_sol}

In this section, we present our main result of the paper. We use the exponential mixing conditions of the network dynamics to estimate the minimum length of time series such that the ergodic basis pursuit has a unique solution. Here we avoid the multi-index notation in $\mathcal{L}$ and $\mathcal{L}_{\nu}$ used in the previous section and instead employ the notation that makes explicit the ordering index as $\phi_l:M^{N} \to \mathbb{R}$ with $l \in [m]$. More precisely, in an explicit form, we say that each $\phi_l$ corresponds to a function of the form $\phi_{ij}^{pq} \circ \pi_{\mathcal{J}}$ for a particular $\mathcal{J} \subset [N]$ with $i, j \in \mathcal{J}$. Also, we use the distance between probability measures introduced in \ref{ssec:metric_prob_measures}. 

\begin{theorem}\label{thm:appendix_noiseless_case}
Let $(F_{\alpha}, \mu_{\alpha})$ be an exponential mixing network dynamical system on $M^N$ with decay exponent $\gamma > 0$ uniform on $N$. Let $\nu = \prod_{i \in [N]} \nu_i \in \mathcal{M}(M^N)$ be a product probability measure and absolutely continuous w.r.t. Lebesgue.  Let $\mathcal{L}_{\nu}$ be the orthonormal network library with respect to $\nu$ and cardinality $m = \binom{N}{2} \binom{r}{2} + N r + 1$. Let $\mathcal{K} = (\mathcal{L}_{\nu} \cdot \mathcal{L}_{\nu})$ and $\omega_r(s)$ satisfies \eqref{eq:omega_r_s}. Suppose that given $\alpha > 0$ there is $\zeta \in (0, \frac{\sqrt{2} - 1}{2\omega_r(s) +\sqrt{2} - 2})$ such that $d_{\mathcal{K}}(\nu, \mu_{\alpha}) < \zeta$ and each one-dimensional marginal $\nu_i$ has Lipschitz density $\rho_i$ with constant $\mathrm{Lip}(\rho) = \max_{i \in [N]}\mathrm{Lip}(\rho_i) $ and $\rho_{0} = \min_{i \in [N]} \{\min_{x \in M_i} \rho_i(x)\} > 0$. Then: 
\begin{enumerate}
    \item \label{thm:phi_nu_rip} \emph{[$\Phi_{\nu}(X)$ satisfies RIP]} Given $\lambda \in (0, 1)$ there exists set of initial conditions $\mathcal{G} \subset M^N$ with probability $\mu_{\alpha}(\mathcal{G}) \geq 1 - \lambda$ if the length of time series $n$ satisfies
\begin{linenomath} \begin{align}\label{eq:minimum_length_time_series}
    n \geq K_1 \frac{(2\omega_r(s) + \sqrt{2}  -2)^2}{\big(\sqrt{2} - 1 - \zeta (2\omega_r(s) + \sqrt{2} - 2)\big)^2}\ln{\Big(\frac{4m(m - 1)}{\lambda}\Big)},
\end{align}\end{linenomath}
for some positive constant $K_1 = K_1(\mathrm{Lip}(\rho), \rho_0)$, then $\Phi_{\nu}(X)$ satisfies the RIP with constant $\delta_{2\omega_r(s)} \leq \sqrt{2} - 1$. 
\item \label{thm:ebp_has_unique_sol} \emph{[EBP has unique solution]} Consider that the length of time series $n$ satisfies  \eqref{eq:minimum_length_time_series}. Let $\bar{x} = \Phi_{\nu}(X)c_{\nu}$ where $c_{\nu} \in \mathbb{R}^m$ is an $\omega_r(s)$-sparse vector and consider the set $
\mathcal{F}_{\bar{x}} = \{w \in \mathbb{R}^{m} ~:~ \Phi_{\nu}(X) w = \bar{x}\}$. Then $c_{\nu}$ is the unique minimizer of the ergodic basis pursuit: 
\begin{linenomath} \begin{align}\label{eq:ergodic_BP}
\begin{split}
\mathrm{(EBP)} \quad  \min_{u \in \mathcal{F}_{\bar{x}}} \|u\|_1.
\end{split}
\end{align}\end{linenomath}
\end{enumerate}
\end{theorem}
The above theorem has an asymptotic expression for sufficient large networks and small $\zeta$ to a simpler condition on the length of time series:
\begin{coro}
For sufficiently large $N > 0$, if the length of time series $n$ satisfies
\begin{linenomath} \begin{align}\label{eq:min_length_time_series_paper}
\begin{split}
n \geq n_0 = \frac{20 K_1 \omega_r^2(s)}{(\sqrt{2} - 1)^2} \ln{(N r)} + \mathcal{O}(\zeta) + \mathcal{O}(\frac{1}{Nr}).
\end{split}
\end{align}\end{linenomath}
then with probability at least $1 - \frac{4}{N r}$ the restricted isometry constant $\delta_{2 \omega_r(s)} \leq \sqrt{2} - 1$.
\end{coro}
\begin{proof}
Assume that \eqref{eq:minimum_length_time_series} holds. Recall that $m = \binom{N}{2} \binom{r}{2} + N r + 1$, then $m < (N r + 1)^2$. Given $\lambda \in (0,1)$ there exists $N_0 > 0$ such that for any $N \geq N_0$: $\frac{4}{Nr} \leq \lambda$. Then, the following holds
\begin{linenomath} \begin{align*}
\ln{\Big(\frac{4m(m - 1)}{\lambda}\Big)} &< \ln{\Big(\frac{4 (Nr + 1)^2}{\lambda}\Big)} \\ 
&= \ln{\Big(\frac{4 (Nr)^4 (1 + \frac{1}{Nr})^4}{\lambda}\Big)} \\
&\leq \ln{(Nr)^5(1 + \frac{1}{Nr})^4} \\
&=\ln{(Nr)^5} + \mathcal{O}(\frac{1}{Nr}).
\end{align*} \end{linenomath} 
Also, for $\zeta \in (0, \frac{\sqrt{2} - 1}{2\omega_r(s) +\sqrt{2} - 2})$, we can expand in geometric series:
\begin{linenomath} \begin{align*}
\frac{1}{(\sqrt{2} - 1 - \zeta (2\omega_r(s) + \sqrt{2} - 2))^2} &= \frac{1}{(\sqrt{2} - 1)^2(1 - \frac{\zeta (2\omega_r(s) + \sqrt{2} - 2)}{\sqrt{2} - 1})^2}\\
&= \frac{1}{(\sqrt{2} - 1)^2} (1 + \mathcal{O}(\zeta)).
\end{align*} \end{linenomath} 
So, we obtain the claim. 
\end{proof}

We split proof of Theorem \ref{thm:appendix_noiseless_case} in steps detailed in the sections below. First, we show that the Bernstein-like inequality applied to $(F_{\alpha}, \mu_{\alpha})$ implies that there exists $n_0$ such that the library matrix $\Phi_{\nu}(X)$ associated to $\nu$ has the desired restricted isometry constant. Then, we apply Theorem \ref{thm:candes_exact_recovery} to demonstrate that the ergodic basis pursuit in  \eqref{eq:ergodic_BP} has a unique solution.

\subsection{Network library matrix satisfies RIP}
\label{sec:net_lib_RIP}
We begin this section by proving an auxiliary lemma that will be used later. 
\begin{lemma}\label{lemma:sup_l2_norm_bounds}
Let $\nu = \prod_{i \in [N]} \nu_i \in \mathcal{M}(M^N)$ be a product probability measure. Suppose that each one-dimensional marginal $\nu_i$ is absolutely continuous w.r.t. Lebesgue and its density is Lipschitz with constant $\mathrm{Lip}(\rho)$ and $\rho_{0} = \min_{i \in [N]} \{\min_{x \in M_i} \rho_i(x)\} > 0$. Let $\mathcal{L}_{\nu}$ be the orthonormal network library and $\mathcal{K} = (\mathcal{L}_{\nu} \cdot \mathcal{L}_{\nu})$. Given $\alpha > 0$ and $\zeta > 0$ sufficiently small, suppose that $d_{\mathcal{K}}(\nu, \mu_{\alpha}) < \zeta$. Denote $(\psi_i \cdot \psi_j) = (\varphi_i \cdot \varphi_j) - \int_{M^N} (\varphi_i \cdot  \varphi_j) d\mu_{\alpha}$. Then, the following holds:
\begin{enumerate}
\item $\max_{i, j}\|(\psi_{i} \cdot \psi_{j})\|_{\infty} \leq 2 \max\{1, K^2\}.
$
\item $\max_{i, j}\|(\psi_{i} \cdot \psi_{j})\|_{\mu_{\alpha}}^2 \leq \max\{1, K^4\} + (1 + \zeta)^2$,
\end{enumerate}
where $K > 0$ is the positive constant in \eqref{eq:constant_K_rho}.
\end{lemma}
\begin{proof}
To prove item 1, note that:
\begin{linenomath} \begin{align*}
\|(\psi_{i} \cdot \psi_{j})\|_{\infty} &\leq \|(\varphi_i \cdot \varphi_j)\|_{\infty} + \int |(\varphi_i \cdot \varphi_j)| d \mu_{\alpha} \\
& \leq  2\|(\varphi_i \cdot \varphi_j)\|_{\infty}.
\end{align*} \end{linenomath} 
To calculate the sup norm of the product of two orthonormal polynomials in $\mathcal{L}_{\nu}$, we consider the notation of the previous section in the following cases:
\begin{linenomath} \begin{align*}
(\varphi_i \cdot \varphi_j) = \begin{cases} 
(1 \cdot 1), \\
(1 \cdot \varphi_{i0}^{p0}), \\
(1 \cdot \varphi_{ij}^{pq}), \\
(\varphi_{i0}^{p0} \cdot \varphi_{jk}^{ql}), \\
(\varphi_{ij}^{pq} \cdot \varphi_{km}^{ln}).
\end{cases}
\end{align*} \end{linenomath} 
By Proposition \ref{propo:one_variable_sup} and Corollary \ref{coro:two_variables_sup},
\begin{linenomath} \begin{align*}
\|(\varphi_i \cdot \varphi_j)\|_{\infty} = \begin{cases}
\max\{K^{\frac{1}{2}}, K, K^{\frac{3}{2}}, K^2\}, \quad i \neq j \\
\max\{1, K, K^2\}, \quad i = j.
\end{cases}
\end{align*} \end{linenomath} 
A priori, the constant $K$ is a given positive number, so it is enough to consider $\|(\psi_{i} \cdot \psi_{j})\|_{\infty} \leq \max\{1, K^2\}$, proving item 1.

To prove item 2, note that given
\begin{linenomath} \begin{align*}
\left|\int_{M^N} (\varphi_i \cdot \varphi_j) d\mu_{\alpha} - \int_{M^N} (\varphi_i \cdot \varphi_j) d\nu \right| \leq d_{\mathcal{K}}(\nu, \mu_{\alpha}) \leq \zeta.
\end{align*} \end{linenomath} 
Consequently, by the triangular inequality
\begin{linenomath} \begin{align*}
\left|\int_{M^N} (\varphi_i \cdot \varphi_j) d\mu_{\alpha}\right| \leq \begin{cases}
1 + \zeta, \quad i = j \\
\zeta, \quad \mathrm{otherwise}. 
\end{cases}
\end{align*} \end{linenomath} 
Then to prove the statement suffices to use item 1 above:
\begin{linenomath} \begin{align*}
\|(\varphi_i \cdot \varphi_j) \|_{\mu_{\alpha}}^2 &= \left|\int (\varphi_i \cdot \varphi_j)^2 d\mu_{\alpha} - \Big(\int \varphi_i^2 d\mu_{\alpha}\Big)^2 \right| \\
&\leq \|(\varphi_i \cdot \varphi_j)\|_{\infty}^2 + (1+\zeta)^2 \\
& \leq \max\{1, K^4\} + (1+\zeta)^2.
\end{align*} \end{linenomath} 
\end{proof}

\subsubsection{Proof of Theorem \ref{thm:appendix_noiseless_case}.\ref{thm:phi_nu_rip}}
\label{sec:proof_n_0}
The following proposition proves that the matrix $\Phi_{\nu}$ attains the desired RIP constant once the length of time series is given by \eqref{eq:n_conditions_delta}.

\begin{propo}\label{propo:appendix_n_min_RIP}
Consider the setting of Theorem \ref{thm:appendix_noiseless_case}. Given $\delta \in (0, 1)$ and $\alpha > 0$, suppose that there is $\zeta \in (0, \frac{\delta}{\delta + \omega_r(s) - 1})$ such that $d_{\mathcal{K}}(\nu, \mu_{\alpha}) < \zeta$.  Then, given $\lambda \in (0, 1)$ there exists a set of initial conditions $\mathcal{G} \subset M^N$ with probability $\mu_{\alpha}(\mathcal{G}) \geq 1 - \lambda$ such that
\begin{linenomath} \begin{align}\label{eq:n_conditions_delta}
    n \geq K_1 \frac{(\delta + \omega_r(s) - 1 )^2}{(\delta - \zeta (\delta + \omega_r(s) - 1))^2}\ln{\Big(\frac{4m(m - 1)}{\lambda}\Big)}
\end{align}\end{linenomath}
for a positive constant $K_1$, then the restricted isometry constant $\delta_{\omega_r(s)}$ of $\Phi_\nu(X)$ satisfies $\delta_{\omega_r(s)} \leq \delta$.
\end{propo}
\begin{proof}
We develop the argument for a coordinate of $F_{\alpha}$. 
Let
\begin{linenomath} \begin{align*}
    u_i :=  
       \frac{1}{\sqrt{n}} \begin{pmatrix}
           \varphi_i(x_0) \\
           \vdots \\
           \varphi_i(F_{\alpha}^{n - 1}(x_0)) \\
        \end{pmatrix} \qquad u_j :=
       \frac{1}{\sqrt{n}} \begin{pmatrix}
           \varphi_j(x_0) \\
           \vdots \\
           \varphi_j(F_{\alpha}^{n - 1}(x_0)) \\
        \end{pmatrix}
\end{align*} \end{linenomath} 
be the $i$-th and $j$-th columns of the matrix $\Phi_{\nu}(X) \in \mathbb{R}^{n \times m}$ for an arbitrary initial condition $x_0 \in M^N$, and their inner product
\begin{linenomath} \begin{align*} 
    \langle u_i, u_j \rangle &= \frac{1}{n} \sum_{k=0}^{n-1} \varphi_i(F_{\alpha}^k(x_0)) \varphi_j(F_{\alpha}^k(x_0))  \\
&= \frac{1}{n} \sum_{k=0}^{n-1} (\varphi_i \cdot  \varphi_j )\circ (F_{\alpha}^k(x_0)) \\
&=: \frac{1}{n} S_n(\varphi_i \cdot  \varphi_j )(x_0). 
\end{align*} \end{linenomath} 
We aim to estimate this inner product using the inner product in $L^2(\nu)$. By triangular inequality, we know that:
\begin{linenomath} \begin{align}\label{eq:triangular_ineq_birkhoff}
\begin{split}
\left|\frac{1}{n}S_n(\varphi_i \cdot \varphi_j)(x_0) - \int_{M^N} (\varphi_i \cdot  \varphi_j) d\nu \right| &\leq \left|\frac{1}{n}S_n(\varphi_i \cdot \varphi_j)(x_0) - \int_{M^N} (\varphi_i \cdot \varphi_j) d\mu_{\alpha} \right| + \\ &\qquad\underbrace{\left|\int_{M^N} (\varphi_i \cdot \varphi_j) d\mu_{\alpha} - \int_{M^N} (\varphi_i \cdot  \varphi_j) d\nu \right|}_{|h_{ij}|}.
\end{split}
\end{align}\end{linenomath}
We introduce a variant of $(\varphi_i \cdot \varphi_j)$ to have zero mean with respect to $\mu_{\alpha}$, i.e., let us denote $(\psi_i \cdot \psi_j) = (\varphi_i \cdot \varphi_j) - \int_{M^N} (\varphi_i \cdot  \varphi_j) d\mu_{\alpha}$ and by hypothesis, 
\begin{linenomath} \begin{align}\label{eq:bound_h_ij}
\begin{split}
|h_{ij}| &= \left|\int_{M^N} (\varphi_i \cdot  \varphi_j) d\mu_{\alpha} - \int_{M^N} (\varphi_i \cdot  \varphi_j) d\nu \right| \\
&\leq d_{\mathcal{K}}(\nu, \mu_{\alpha}) \\
&\leq \zeta.
\end{split}
\end{align}\end{linenomath}
%In \eqref{eq:triangular_ineq_birkhoff}, fix $\eta_0 > 0$ such that 
%\begin{linenomath} \begin{align*}
%\left|\frac{1}{n}S_n(\varphi_i \cdot \varphi_j)(x_0)\right| \geq \eta_0,
%\end{align*} \end{linenomath} 
%consequently, 
Then, we split into two distinct cases that run in parallel:
\begin{enumerate}
\item \emph{$i \neq j$}: $ \int_{M^N} (\varphi_i \cdot  \varphi_j) d\nu = 0$, consequently, using \eqref{eq:bound_h_ij} we conclude that follows
\begin{linenomath} \begin{align}\label{eq:S_n_varphi_i_j}
\left|\frac{1}{n}S_n(\varphi_i \cdot \varphi_j)(x_0)\right| \leq \left|\frac{1}{n}S_n(\psi_i \cdot \psi_j)(x_0)\right| + |h_{ij}| \leq  \left|\frac{1}{n}S_n(\psi_i \cdot \psi_j)(x_0)\right| + \zeta.
\end{align}\end{linenomath}
\item \emph{$i = j$}: we have $\int_{M^N}\varphi_i^2 d\nu = 1$, and consequently, in \eqref{eq:triangular_ineq_birkhoff}, we obtain
\begin{linenomath} \begin{align*}
\left|\frac{1}{n}S_n(\varphi_i^2)(x_0) - 1\right| \leq \left|\frac{1}{n}S_n(\psi_i^2)(x_0)\right| + \zeta.
\end{align*} \end{linenomath} 
By the triangular inequality, we conclude that
\begin{linenomath} \begin{align}\label{eq:S_n_varphi_i_2}
\left|\frac{1}{n}S_n(\varphi_i^2)(x_0)\right| \geq 1 - \left|\frac{1}{n}S_n(\psi_i^2)(x_0)\right| - \zeta.
\end{align}\end{linenomath}
\end{enumerate}
Note that $(\psi_i \cdot  \psi_j)$ and $\psi_i^2$ are given by a finite linear combination of elements in $\mathcal{L}$, and consequently, the subsets with cardinality $\binom{m}{2}$ and $m$, respectively, satisfy
\begin{linenomath} \begin{align}\label{eq:set_of_observables_1}
\mathcal{K}_1 = \{ (\psi_{i}\cdot \psi_j) ~:~i, j = 1, \dots, m, i \neq j\} \subset \mathcal{C}^1(M^N;\mathbb{R})
\end{align}\end{linenomath}
and
\begin{linenomath} \begin{align}\label{eq:set_of_observables_2}
\mathcal{K}_2 = \{ \psi_{i}^2 ~:~i = 1, \dots, m\} \subset \mathcal{C}^1(M^N;\mathbb{R}).
\end{align}\end{linenomath}
Choose $\varkappa > 0$, $\varsigma > 0$ and $\sigma > 0$ such that 
\begin{linenomath} \begin{align}\label{eq:constants_bernstein}
\begin{split}
\varkappa &:= \max \{\max_{i \neq j} \|(\psi_i \cdot \psi_j)\|_{\infty}, \max_{i \in [m]} \|\psi_i^2\|_{\infty}\}, \\
\varsigma &:= \max\{\max_{i \neq j} \|D(\psi_i \cdot \psi_j)\|_{\infty}, \max_{i \in [m]} \|D\psi_i^2\|_{\infty}\}, \\
\sigma^2 &:= \max \{\max_{i \neq j} \|(\psi_i \cdot \psi_j)\|_{\mu_{\alpha}}^2, \max_{i \in [m]} \|\psi_i^2\|_{\mu_{\alpha}}^2\}.
\end{split}
\end{align}\end{linenomath}
By the Bernstein inequality in Theorem \ref{thm:bernstein_inequality}, for $\eta_0 > 0$ and $n \geq n_0(\varkappa, \varsigma, \sigma, \gamma)$, which is defined in  \eqref{eq:n_0_bound}, if we define
\begin{linenomath} \begin{align*}
\mathcal{O}_1 &= \bigcup_{i \neq j} \Big\{x_0 \in M^N ~:~\left|\frac{1}{n}S_n(\psi_i \cdot \psi_j)(x_0) \right| \geq \eta_0 \Big\} \\ 
\mathcal{O}_2 &= \bigcup_{i \in [m]} \Big\{x_0 \in M^N ~:~\left|\frac{1}{n}S_n(\psi_i^2)(x_0) \right| \geq \eta_0 \Big\},
\end{align*} \end{linenomath} 
then 
\begin{linenomath} \begin{align*}
    \mu_{\alpha}(\mathcal{O}_1) \leq  4 \binom{m}{2} e^{-\theta(\eta_0, n, \sigma, \varkappa)} \quad \mathrm{and} \quad     \mu_{\alpha}(\mathcal{O}_2) \leq  4 m e^{-\theta(\eta_0, n, \sigma, \varkappa)}.
\end{align*} \end{linenomath} 
We are interested in the case $\mu_{\alpha}(\mathcal{O}) = \mu_{\alpha}(\mathcal{O}_1 \cup \mathcal{O}_2)$
\begin{linenomath} \begin{align*}
\mu_{\alpha}(\mathcal{O}) &\leq 4 \binom{m}{2} e^{-\theta(\eta_0, n, \sigma, \varkappa)} + 4 m e^{-\theta(\eta_0, n, \sigma, \varkappa)} \\
&\leq 8 \binom{m}{2} 
e^{-\theta(\eta_0, n, \sigma, \varkappa)}.
\end{align*} \end{linenomath} 
%&\leq 8 {m \choose 2} e^{-\theta(\eta_1, n, \sigma, \varkappa)}, 
%
%where we use that $\theta(\eta_1, \cdot) < \theta(\eta_1 + 1, \cdot)$ for any positive $\eta_1 > 0$. In fact, $\eta_1 + 1$ is a translation to the left in the graph of the function, which is strictly increasing for positive $\eta_1$.
For the given $\lambda \in (0, 1)$ the set $\mathcal{O}^c \subset M^{N}$ of initial conditions, whose Birkhoff sum satisfies the desired precision $\eta_1$, has measure $\mu_{\alpha}(\mathcal{O}^c) \geq 1 - \lambda$ whenever
\begin{linenomath} \begin{align}\label{eq:n_1_bounds}
    \frac{n}{(\ln n)^{2}} \geq \frac{8}{\eta_0^2} (\sigma^2 + \varkappa \frac{\eta_0}{3}) \ln{\Big(\frac{8}{\lambda}\binom{m}{2}\Big)}.
\end{align}\end{linenomath}
Instead of  \eqref{eq:n_1_bounds}, one usually prefers a condition that features only $n$ on the left-hand side. First, note that whenever $n \geq n_0$ implies that $n \in \mathcal{N}$ in  \eqref{eq:n_0_bound}, and consequently, the function $t \mapsto t/(\ln t)^2$ is monotonic for values in $\mathcal{N}$. So, the condition in \eqref{eq:n_1_bounds} is in fact implied by
\begin{linenomath} \begin{align}\label{eq:proof_n_bounds}
n \geq \frac{8}{\eta_0^2} (\sigma^2 + \varkappa \frac{\eta_0}{3}) \ln{\Big(\frac{4 m (m - 1)}{\lambda}\Big)}.
\end{align}\end{linenomath}
For any $n$ satisfying the bound in  \eqref{eq:proof_n_bounds}, we can normalize any two distinct columns vectors $u_i$ and $u_j$, which we denote $v_i$ and $v_j$, respectively. So, we can estimate the coherence of the matrix $\Phi_{\nu}(X)$ for any $x_0 \in \mathcal{O}^c$ using \eqref{eq:S_n_varphi_i_j} and \eqref{eq:S_n_varphi_i_2}
\begin{linenomath} \begin{align*}
\eta(\Phi_{\nu}) &:= \max_{i \neq j} |\langle v_i, v_j\rangle| =  \max_{i \neq j} \frac{|\langle u_i, u_j\rangle|}{\|u_i\|_2 \|u_j\|_2} \\ 
&= \max_{i \neq j} \frac{\Bigl|\frac{1}{n} S_n (\varphi_i \cdot \varphi_j)  (x_0)\Bigr|}{\Bigl|\frac{1}{n} S_n (\varphi_i^2)(x_0)\Bigr|^{\frac{1}{2}}\Bigl|\frac{1}{n} S_n (\varphi_j^2)(x_0)\Bigr|^{\frac{1}{2}}}  \\
&\leq \frac{\eta_0 + \zeta}{1 - (\eta_0 + \zeta)},
\end{align*} \end{linenomath} 
which is valid for a $\eta_0$ such that $\eta_0 + \zeta < 1$. Finally, the desired restricted isometry constant is attained because the coherence upper bounds the restricted isometry constant of the matrix $\Phi_{\nu}(X)$ by Proposition \ref{prop:coherence_upper_bounds}. So, for the given $\delta \in (0, 1)$, we choose 
\begin{linenomath} \begin{align}\label{eq:eta_0_expression}
\eta_0 = \frac{\delta - \zeta(\delta + \omega_r(s) - 1)}{\delta + \omega_r(s) - 1} 
\end{align}\end{linenomath}
that is positive as long as $\zeta \in (0, \frac{\delta}{\delta + \omega_r(s) - 1})$. 
To obtain the desired bounds on the length of the time series, note that $\eta_0 \in (0, 1)$. Also, we use Lemma \ref{lemma:sup_l2_norm_bounds} in order to bound $\sigma^2$ and $\varkappa$ in \eqref{eq:constants_bernstein}. Consequently, the condition in \eqref{eq:n_1_bounds} is also implied by
\begin{linenomath} \begin{align*}
n \geq \frac{8}{\eta_0^2} (\max\{1, K^4\} + (1 + \zeta)^2 +  \frac{2}{3}\max\{1, K^2\}) \ln{\Big(\frac{4 m (m - 1)}{\lambda}\Big)},
\end{align*} \end{linenomath} 
which can also be implied by
\begin{linenomath} \begin{align}\label{eq:proof_n_bounds_1}
n \geq \frac{K_1}{\eta_0^2}\ln{\Big(\frac{4 m (m - 1)}{\lambda}\Big)},
\end{align}\end{linenomath}
with $K_1 := 8(\max\{1, K^4\} + 4 +  \frac{2}{3}\max\{1, K^2\})$. Replacing \eqref{eq:eta_0_expression} in \eqref{eq:proof_n_bounds_1}, we obtain the result.
\end{proof}

\begin{lemma}\label{lemma:uniqueness_c_statement}
Let $c_{\nu} \in \mathbb{R}^{m}$ be a $\omega_r(s)-$sparse vector. If the length of time series $n$ satisfies  \eqref{eq:minimum_length_time_series}, then EBP in  \eqref{eq:ergodic_BP} has $c_{\nu}$ as its unique solution. 
\end{lemma}
\begin{proof}
Combining Proposition \ref{propo:appendix_n_min_RIP} with Theorem \ref{thm:candes_exact_recovery} suffices.
\end{proof}

\subsection{Ergodic basis pursuit has a sufficient infeasibility condition }

Since Theorem \ref{thm:appendix_noiseless_case} ensures that EBP has a unique solution, we can also prove an additional result.

\begin{propo}[Sufficient infeasibility condition]\label{thm:infeasibility} 
Consider that the length of time series $n$ satisfies  \eqref{eq:minimum_length_time_series}. Let $\bar{x} = \Phi_{\nu}(X)c_{\nu}$ where $c_{\nu} \in \mathbb{R}^m$ is an $\omega_r(s)$-sparse vector and consider the set $\mathcal{F}_{\bar{x}} = \{w \in \mathbb{R}^{m}: \Phi_{\nu}(X) w = \bar{x}\}$. 
Given a set $\mathcal{U} \subseteq [m]$ where $\mathcal{U} \cap \mathrm{supp}(c_{\nu}) \neq \emptyset$. Then $\mathcal{U} \subsetneqq \mathrm{supp}(c_{\nu})$ if and only if
\begin{linenomath} \begin{align}\label{eq:infeasibility_condition}
\mathcal{F}_{\bar{x}} \bigcap \{w \in \mathbb{R}^{m}~:~\mathrm{supp}(w) = \mathcal{U}\} =
\emptyset.
\end{align}\end{linenomath}
\end{propo}
\begin{proof}
Let us assume that $\mathcal{I} \subsetneqq \mathrm{supp}(c_{\nu})$. We will prove this by contradiction. Suppose there is a vector $w \neq 0$ in the intersection  \eqref{eq:infeasibility_condition} and is given by 
\begin{linenomath} \begin{align*}
w = (w_1, \dots, w_{\omega_r(s) - 1}, 0, \dots, 0)
\end{align*} \end{linenomath} 
as opposed to the $\omega_r(s)$-sparse vector $c$,
\begin{linenomath} \begin{align*}
c_{\nu} = (c_1, \dots, c_{\omega_r(s)}, 0, \dots, 0),
\end{align*} \end{linenomath} 
so, the vector $w$ has $\omega_r(s)-1$ nonzero entries. Since $w \in \mathcal{F}_{\bar{x}}$, we have
\begin{linenomath} \begin{align*}
\Phi_{\nu}(X)w = \Phi_{\nu}(X) c_{\nu}.
\end{align*} \end{linenomath} 
Consequently, 
\begin{linenomath} \begin{align*}
\Phi_{\nu}(X)(w - c_{\nu}) = 0.
\end{align*} \end{linenomath} 
But the $\Phi_{\nu}(X)$ satisfies RIP with constant $\delta_{2\omega_r(s)} < \sqrt{2} - 1$. Since $w - c_{\nu}$ is an $\omega_r(s)$-sparse vector, we calculate
\begin{linenomath} \begin{align*}
\|\Phi_{\nu}(X)(w - c_{\nu})\|_2^2 \geq (1 - \delta_{2\omega_r(s)}) \|w - c_{\nu}\|_2^2 > 0.
\end{align*} \end{linenomath} 
So, we conclude that this is only possible when $w = c$, which is a contradiction since $c$ is not in the intersection, and the claim follows. 

The other direction we prove by contrapositive. We contradict $\mathcal{I} \subsetneqq  \mathrm{supp}(c_{\nu})$. Since $\mathcal{I}$ must have an intersection with $\mathrm{supp}(c_{\nu})$, then suppose that $\mathrm{supp}(c_{\nu}) \subseteq \mathcal{I}$. The intersection \eqref{eq:infeasibility_condition} is non-empty, since the sparse vector $c_{\nu}$ is an element of the set. This proves the claim, and the statement follows.
\end{proof}

\section{Noise measurement case}
\label{sec:noise_robust_reconstruction}

Here, we extend the Ergodic Basis Pursuit to reconstruct the network from corrupted measurements
\begin{linenomath} \begin{align}\label{eq:appendix_noisy_data}
    y(t) = x(t) + z(t),
\end{align}\end{linenomath}
such that $(z_n)_{n \geq 0}$ corresponds to independent and identically distributed $[-\xi, \xi]^{N}$-valued noise process for $\xi \in (0, 1)$ with probability measure $\varrho_{\xi}$. Let the convolution $\mu_{\alpha, \xi} = \mu_{\alpha} * \varrho_{\xi}$ be the probability measure of the process $(y_n)_{n \geq 0}$ \cite{folland2013real} and the matrix $\bar{Y}$ be the noisy data
\begin{linenomath} \begin{align}\label{eq:appendix_noisy_meas_matrix}
\bar{Y} = \left(
\begin{array}{ccc}
y_1(1) & \cdots &  y_N(1)	    \\
\vdots  & \ddots &  \vdots\\
y_1(n)   &  \cdots &  y_N(n)	  
\end{array} \right).
\end{align}\end{linenomath}
The next theorem assumes that there exists a product measure $\nu_{\xi}$ sufficiently close to the measure $\mu_{\alpha, \xi}$. Thus, the $s$-sparse vector $c \in \mathbb{R}^{m}$ corresponding to the representation in $\mathcal{L}$ is mapped to $c_{\nu_{\xi}}$ which represents the network dynamics in $\mathcal{L}_{\nu_{\xi}}$, i.e., it satisfies $\bar{x} = \Phi_{\nu_{\xi}}(X) c_{\nu_{\xi}}$. Here, we also introduce another convex minimization problem in terms of $\Phi_{\nu_{\xi}}(Y)$ evaluated along the process $(y_n)_{n \geq 0}$. We show that the family of solutions of this minimization problem is parametrized by the noise level in such way that approximates the sparse vector $c_{\nu_{\xi}}$.

Here, we rewrite the interval bounds: for a given $i \in [N]$ let $M_{i, \xi} = [a - \xi, b + \xi] \subset \mathbb{R}$ with $b > a$ and $\xi \in (0, 1)$. Then, consider the following:

\begin{hypothesis}\label{hypo:noise_magnitude_small}
Let $\nu_{\xi} = \prod_{i \in [N]} \nu_{i,\xi} \in \mathcal{M}(M^N + [-\xi, \xi]^N)$ be a product probability measure and absolutely continuous w.r.t. Lebesgue. Suppose that given a sufficiently small $\xi > 0$, each one-dimensional marginal $\nu_{i, \xi}$ has Lipschitz density $\rho_{i, \xi}$ with constant $\mathrm{Lip}(\rho_{\xi}) = \max_{i \in [N]}\mathrm{Lip}(\rho_{i, \xi})$ and $\rho_{0, \xi} = \min_{i \in [N]} \{\min_{x \in M_{i, \xi}} \rho_{i, \xi}(x)\} > 0$. 
\end{hypothesis}

\begin{theorem}[Noise reconstruction case]\label{thm:appendix_noise_reconstr}
Consider the setting of Theorem \ref{thm:appendix_noiseless_case} and Hypothesis \ref{hypo:noise_magnitude_small}. Let $\mathcal{L}_{\nu_{\xi}} = \{\varphi_l\}_{l = 1}^{m}$ be the orthonormal ordered network library with respect to $\nu_{\xi}$ and $\mathcal{K} = (\mathcal{L}_{\nu_{\xi}} \cdot \mathcal{L}_{\nu_{\xi}})$. Suppose that given $\alpha > 0$ there is $\zeta \in (0, \frac{\sqrt{2} - 1}{2 \omega_r(s) +\sqrt{2} - 2} - K_1 r^2 \xi)$ such that $d_{\mathcal{K}}(\nu_{\xi}, \mu_{\alpha, \xi}) < \zeta$ for a positive constant $K_1 = K_1(\mathrm{Lip}(\rho_{\xi}), \rho_{0, \xi}, \xi)$. Then
\begin{enumerate}
\item \label{thm:noisy_net_lib_RIP} \emph{[$\Phi_{\nu}(Y)$ satisfies RIP]} Given $\lambda \in (0, 1)$ there exists a set of initial conditions $\mathcal{G} \subset M^N$ with probability $\mu_{\alpha}(\mathcal{G}) \geq 1 - \lambda$ such that if the length of time series $n$ satisfies
\begin{linenomath} \begin{align}\label{eq:n_condition_noisy}
   n \geq K_2 \frac{(2\omega_r(s) + \sqrt{2} - 2)^2}{(\sqrt{2} - 1 - (\zeta + K_1 r^2 \xi) (2\omega_r(s) + \sqrt{2} - 2))^2}\ln{\Big(\frac{4m(m - 1)}{\lambda}\Big)},
\end{align}\end{linenomath}
for a positive constant $K_2 = K_2(\mathrm{Lip}(\rho_{\xi}), \rho_{0, \xi}, \xi)$, then the restricted isometry constant $\delta_{2\omega_r(s)}$ of $\Phi_{\nu_{\xi}}(X)$ satisfies $\delta_{2\omega_r(s)} \leq \sqrt{2} - 1$.
\item \label{thm:EBP_is_robust} \emph{[EBP is robust]} Let $\bar{y} \in M^n + [-\xi, \xi]^n$ be a column of $\bar{Y}$, $c_{\nu_{\xi}} \in \mathbb{R}^{m}$ be an $\omega_r(s)$-sparse vector with $\|c_{\nu_{\xi}}\|_{\infty} < \infty$ such that $\bar{x} = \Phi_{\nu_{\xi}}(X) c_{\nu_{\xi}}$. Consider that the length of time series $n$ satisfies  \eqref{eq:n_condition_noisy}. Then the family of solutions $\{c^{\star}(\epsilon)\}_{\epsilon > 0}$ to the convex problem
\begin{linenomath} \begin{align}\label{eq:QBP_EBP_1}
\begin{split}
\min_{\tilde{u} \in \mathbb{R}^{m}} \|\tilde{u}\|_1 ~\mathrm{subject~to}~ \|\Phi_{\nu_{\xi}}(Y) \tilde{u} - \bar{y}\|_2 \leq  \epsilon
\end{split}
\end{align}\end{linenomath}
satisfies
\begin{linenomath} \begin{align}\label{eq:ell_2_close_solution_1}
    \|c^{\star}(\epsilon) - c_{\nu_{\xi}}\|_2 \leq K_3 \epsilon
\end{align}\end{linenomath}
as long as 
\begin{linenomath} \begin{align}\label{eq:bound_eps_1}
\epsilon \geq \sqrt{n} \xi \Big(1 + m N r^2 K_4 \|c_{\nu_{\xi}}\|_{\infty} \Big),
\end{align}\end{linenomath} 
for positive constants $K_3 = K_3(\delta_{2\omega_s(r)})$ and $K_4 = K_4(\mathrm{Lip}(\rho_{\xi}), \rho_{0, \xi})$.
\end{enumerate}
\end{theorem}

We prove the above theorem in steps detailed in the sections below. First, we adapt the estimate of the minimum length of time series such that the library matrix has the desired restricted isometry constant. Subsequently, we show that the unique solution of the ergodic basis pursuit in  \eqref{eq:ergodic_BP} is approximated in $\ell_2$ by a family of solutions $\{c^{\star}(\epsilon)\}_{\epsilon \geq 0}$.

\subsection{Perturbed network library matrix satisfies RIP}

We begin estimating the distance between the product measure $\nu_{\xi}$ and the physical measure $\mu_{\alpha}$ of the deterministic network dynamics. To this end, we use the auxiliary lemmas below. First, we show that
\begin{lemma}\label{lemma:weak_convergence_measure_mu_xi}
$\mu_{\alpha, \xi} \to \mu_{\alpha}$ converges weakly as $\xi \to 0$.
\end{lemma}
\begin{proof}
Fix a continuous function $\varphi:M^N \to \mathbb{R}$ and $\xi > 0$. Using the definition $\int \varphi d \mu_{\alpha, \xi} = \int \varphi d \mu_{\alpha} * \varrho_{\xi} = \int  \int \varphi(x + z) d \mu_{\alpha}(x) d\varrho_{\xi}(z)$ and $\int \varphi d\mu_{\alpha} = \int \int \varphi(x) d \mu_{\alpha}(x) d\varrho_{\xi}(z)$, we obtain
\begin{linenomath} \begin{align*}
\Big| \int \varphi d \mu_{\alpha, \xi} -  \int \varphi d \mu_{\alpha} \Big| &= \Big| \int \int \varphi(x + z) d \mu_{\alpha}(x) d \varrho_{\xi}(z) -  \int \int \varphi(x) d \mu_{\alpha}(x) d \rho_{\xi}(z)\Big|\\
&\leq \int \Big( \int |\varphi(x + z) - \varphi(x)| d \varrho_{\xi}(z)  \Big)d \mu_{\alpha}(x) \\
&\leq \int \sup_{|z| \leq \xi} |\varphi(x + z) - \varphi(x)| d \mu_{\alpha}(x).
\end{align*} \end{linenomath} 
Since $M^N$ is a compact set, $\varphi$ is uniformly continuous. Then, letting $\xi \to 0$ implies that the right-hand side converges to zero, and consequently, the integrals in the left-hand side converge. This is valid for any continuous function $\varphi$, concluding the statement.
\end{proof}
We address to estimate the distance $d_{\mathcal{K}}(\nu, \mu_{\alpha})$. Since the product measure $\nu_{\xi}$ is defined on $(M^N + [-\xi, \xi]^N)$. Also, we define a variant of the constant \eqref{eq:constant_K_rho} given by
\begin{linenomath} \begin{align}\label{eq:constant_K_rho_xi}
K_{\xi} = K(\mathrm{Lip}(\rho_{\xi}), \rho_{0, \xi}, \xi) \equiv \Big( \frac{1}{\rho_{0, \xi}} + 2\frac{a_1 \mathrm{Lip}(\rho)}{\rho_{0, \xi}^{3/2}}\Big)^2 \Big(\frac{2}{b - a + 2 \xi}\Big)
\end{align}\end{linenomath}
that satisfies $K_{\xi} \to K$ when $\xi \to 0$. Then, the following holds
\begin{lemma}\label{lemma:d_nu_mu_alpha}
Let $r \geq 2$. Given $\alpha, \zeta, \xi \in (0, 1)$ suppose that $d_{\mathcal{K}}(\nu, \mu_{\alpha,\xi}) < \zeta$ for the product measure $\nu_{\xi} \in \mathcal{M}(M^N + [-\xi, \xi]^N)$ . Then, there exists $K_1 = K_1(\mathrm{Lip}(\rho_{\xi}), \rho_{0, \xi}, \xi)$ such that
\begin{linenomath} \begin{align*}
d_{\mathcal{K}}(\nu, \mu_{\alpha}) \leq \zeta + K_1 r^2 \xi.
\end{align*} \end{linenomath} 
\end{lemma}
\begin{proof}
First we calculate $d_{\mathcal{K}}(\mu_{\alpha,\xi}, \mu_{\alpha})$. Fix $\mathcal{J} \subset [N]$ and $\psi \in \mathcal{K}$. Since the projection $\pi_{\mathcal{J}}:M^N \to \prod_{i \in \mathcal{J}} M_i$ is Lipschitz with constant $1$ and $\mathcal{K} $ is a set of product of polynomials, the composition $\psi \circ \pi_{\mathcal{J}}$  is also Lipschitz with constant $\mathrm{Lip}(\psi \circ \pi_{\mathcal{J}}) = \|D \psi\|_{\infty}$. Then, we obtain
\begin{linenomath} \begin{align*}
\left| \int_{M^N} \psi \circ \pi_{\mathcal{J}} d\mu_{\alpha,\xi} - \int_{M^N} \psi \circ \pi_{\mathcal{J}} d \mu_{\alpha} \right| &\leq \int \sup_{|z| \leq \xi} \left|\psi \circ \pi_{\mathcal{J}}(x+z) - \psi \circ \pi_{\mathcal{J}}(x)\right| d \mu_{\alpha}(x) \\
&\leq \mathrm{Lip}(\psi \circ \pi_{\mathcal{J}}) \xi.
\end{align*} \end{linenomath} 
For each $\psi \in \mathcal{K}$ it corresponds to a pair $(\varphi_i \cdot \varphi_j)$, so we use Proposition \ref{propo:one_variable_sup} and Corollary \ref{coro:two_variables_sup} to calculate
\begin{linenomath} \begin{align*}
\mathop{\max_{\mathcal{J} \subset [N]}}_{ 1 \leq |\mathcal{J}| \leq 4} \max_{(\varphi_i \cdot \varphi_j) \in \mathcal{K}} \mathrm{Lip}((\varphi_i \cdot \varphi_j) \circ \pi_{\mathcal{J}}) &\leq \mathop{\max_{\mathcal{J} \subset [N]}}_{ 1 \leq |\mathcal{J}| \leq 4} \max_{(\varphi_i \cdot \varphi_j) \in \mathcal{K}} \|D \varphi_i\|_{\infty} \|\varphi_j\|_{\infty} + \|\varphi_i\|_{\infty} \|D \varphi_j\|_{\infty}
\end{align*} \end{linenomath} 
that is upper bounded by
\begin{linenomath} \begin{align*}
\underbrace{K_{\xi}\max\Big\{\frac{4}{b - a + 2 \xi}, 4 K_{\xi} \Big(\frac{2}{b - a + 2 \xi}\Big)^{\frac{1}{2}}, 2 K_{\xi}^{\frac{1}{2}}\Big(\frac{2}{b - a + 2 \xi}\Big)^{\frac{1}{2}} + \Big(\frac{2}{b - a + 2 \xi}\Big)K_{\xi}^{\frac{1}{2}}\Big\}}_{K_1(\mathrm{Lip}(\rho), \rho_{0, \xi}, \xi)} r^2.
\end{align*} \end{linenomath} 
This yields $d_{\mathcal{K}}(\mu_{\alpha,\xi}, \mu_{\alpha}) \leq  K_1 r^2 \xi$. Using the triangular inequality
\begin{linenomath} \begin{align*}
d_{\mathcal{K}}(\nu, \mu_{\alpha}) \leq d_{\mathcal{K}}(\nu, \mu_{\alpha,\xi}) + d_{\mathcal{K}}( \mu_{\alpha,\xi}, \mu_{\alpha}),
\end{align*} \end{linenomath} 
we conclude the lemma.
\end{proof}
Before we proceed, we extend $\mu_{\alpha} \in \mathcal{M}(M^N)$ to $\mathcal{M}(M^N + [-\xi, \xi]^{N})$, defining the measure of a set $E \subseteq M^N + [-\xi, \xi]^{N}$ as $\mu_{\alpha}(E \cap M^N)$. We abuse notation and denote the measure as $\mu_{\alpha}$.

We can state a similar version of Proposition \ref{propo:appendix_n_min_RIP}, making the appropriate changes. See below:
\begin{propo}\label{propo:appendix_n_min_RIP_noise}
Consider the setting of Theorem \ref{thm:appendix_noise_reconstr}. Given $\delta, \lambda \in (0, 1)$ there exists  a set of initial conditions $\mathcal{G} \subset M^N$ with probability $\mu_{\alpha}(\mathcal{G}) \geq 1 - \lambda$ such that
\begin{linenomath} \begin{align}\label{eq:n_conditions_delta_xi}
    n \geq K_2 \frac{(\omega_r(s) + \delta - 1 )^2}{(\delta - (\zeta + K_1 r^2 \xi) (\omega_r(s) +  \delta - 1))^2}\ln{\Big(\frac{4m(m - 1)}{\lambda}\Big)}
\end{align}\end{linenomath}
for positive constants $K_1$ and $K_2$, then the restricted isometry constant $\delta_{\omega_r(s)}$ of $\Phi_\nu(X)$ satisfies $\delta_{\omega_r(s)} \leq \delta$.
\end{propo}
\begin{proof}
The proof is similar to the proof of Proposition \ref{propo:appendix_n_min_RIP}. Using Lemma \ref{lemma:d_nu_mu_alpha} for the measures $\nu_{\xi}, \mu_{\alpha}$, there is a constant $K_1(\mathrm{Lip}(\rho_{\xi}), \rho_{0, \xi}, \xi)$ such that $d_{\mathcal{K}}(\nu_{\xi}, \mu_{\alpha}) \leq \zeta + K_1 r^2 \xi =: \zeta^{\prime}$, which we define so we can repeat the proof of Proposition \ref{propo:appendix_n_min_RIP} replacing $\zeta$ by $\zeta^{\prime}$. The new bounds of $n_0$ can be deduced as follows: we estimate a new condition that is implied by
\begin{linenomath} \begin{align*}
n \geq \frac{8}{\eta_0^2} (\max\{1, K_{\xi}^4\} + (1 + \zeta + K_1 r^2 \xi)^2 +  \frac{2}{3}\max\{1, K_{\xi}^2\}) \ln{\Big(\frac{4 m (m - 1)}{\lambda}\Big)}.
\end{align*} \end{linenomath} 
This expression can also be implied by
\begin{linenomath} \begin{align}\label{eq:proof_n_bounds_2}
n \geq \frac{K_2}{\eta_0^2}\ln{\Big(\frac{4 m (m - 1)}{\lambda}\Big)},
\end{align}\end{linenomath}
with $K_2 := 8(\max\{1, K_{\xi}^4\} + (2 + K_1 r^2)+ \frac{2}{3}\max\{1, K_{\xi}^2\})$. Using \eqref{eq:eta_0_expression} replacing $\zeta$ by $\zeta+ K_1r^2 \xi$ in the above expression, we obtain the result.
\end{proof}

\begin{proof}[Proof of Theorem \ref{thm:appendix_noise_reconstr}.\ref{thm:noisy_net_lib_RIP}]
It suffices to use Proposition \ref{propo:appendix_n_min_RIP} for $\delta = \sqrt{2} - 1$ and sparsity level $2\omega_r(s)$ in the expression of the length of time series in  \eqref{eq:n_condition_noisy}.
\end{proof}

\subsection{Ergodic basis pursuit is robust against noise}

We can write that $\bar{X} = \Phi_{\nu_{\xi}}(X) C_{\nu_{\xi}}$, where $C_{\nu_{\xi}} \in \mathbb{R}^{m \times N}$ is the coefficient matrix associated to the network library $\mathcal{L}_{\nu_{\xi}}$. We deduce that the noisy data in  \eqref{eq:appendix_noisy_meas_matrix} satisfies 
\begin{linenomath} \begin{align}\label{eq:Y_prime}
\bar{Y} = \Phi_{\nu_{\xi}}(X) C_{\nu_{\xi}} + \bar{Z},
\end{align}\end{linenomath}
where \begin{linenomath} \begin{align}\label{eq:noise_matrix}
\bar{Z} &= \left(
\begin{array}{ccc}
z_1(1) & \cdots &  z_N(1)	    \\
\vdots  & \ddots &  \vdots		\\
z_1(n)   &  \cdots &  z_N(n)	  
\end{array} \right) \in [-\xi, \xi]^{n \times N},
\end{align}\end{linenomath}
such that each column $\bar{z}$ of $\bar{Z}$ is bounded as $\|\bar{z}\|_2 \leq \sqrt{n} \xi$. The following lemma states that the library matrix can be evaluated at the noisy data:
\begin{lemma}\label{lemma:appendix_change_of_data}
\begin{linenomath} \begin{align}\label{eq:phi_mu_Y}
    \Phi_{\nu_{\xi}}(Y) = \Phi_{\nu_{\xi}}(X) + \Lambda(X, \bar{Z}),
\end{align}\end{linenomath}
where $\|\Lambda(X, \bar{Z})\|_{\infty} \leq m N r^2 K_4 \xi$ with $K_4 := \max \{K_{\xi}^{\frac{1}{2}}, 2 K_{\xi} \Big(\frac{2}{b - a + 2\xi}\Big)\}$. 
\end{lemma}
\begin{proof}
For $l \in [m]$ let $\varphi_l \in \mathcal{L}_{\nu_{\xi}}$.  The Mean Value Theorem states that for each $t = 0, \dots, n-1$:
\begin{linenomath} \begin{align*}
    \varphi_l(x(t) + z(t)) = \varphi_l(x(t)) + \left(\int_{0}^{1} D\varphi_l(x(t) + sz(t))ds\right)\cdot z(t),
\end{align*} \end{linenomath} 
where the integral is understood component-wise. Repeating the calculation for each entry of $\Phi_{\nu_{\xi}}(Y)$, by linearity we obtain $\Phi_{\nu_{\xi}}(Y) = \Phi_{\nu_{\xi}}(X) + \Lambda(X, \bar{Z})$, where $\Lambda(X, \bar{Z})$ is the matrix with entries
\begin{linenomath} \begin{align*}
\Lambda_{j,k}(X, Z) = \left( \int_{0}^{1} D\varphi_k(x(j) + sz(j))ds\right)\cdot z(j).
\end{align*} \end{linenomath} 
%{
%\scriptsize	
%\begin{linenomath} \begin{align*}
%\Lambda(X, Z) = \left( \begin{array}{ccc}
%\left(\int_{0}^{1} D\varphi_1(x(0) + sz(0))ds\right)\cdot z(0) & \cdots &  \left(\int_{0}^{1} D\varphi_m(x(0) + sz(0))ds\right)\cdot z(0)	    \\
%\vdots  & \ddots &  \vdots\\
%\left(\int_{0}^{1} D\varphi_1(x(n-1) + sz(n-1))ds\right)\cdot z(n-1)   &  \cdots &  \left(\int_{0}^{1} D\varphi_m(x(n-1) + sz(n-1))ds\right)\cdot z(n-1)	  
%\end{array} \right).
%\end{align*} \end{linenomath} 
%}
We use Proposition \ref{propo:one_variable_sup} and Corollary \ref{coro:two_variables_sup} for $M^{N} + [-\xi, \xi]^{N}$. Let us denote
\begin{linenomath} \begin{align*}
\max_{l \in [m]} \|D \varphi_l\|_{\infty} \leq r^2 \max \{K_{\xi}^{\frac{1}{2}}, 2 K_{\xi} \Big(\frac{2}{b - a + 2\xi}\Big)\} \equiv r^2 K_4.
\end{align*} \end{linenomath} 
Using Cauchy-Schwarz inequality, note that each entry $\Lambda_{j,k}$ satisfies
\begin{linenomath} \begin{align*}
    |\Lambda_{j, k}| &= |\left(\int_{0}^{1} D\varphi_k(x(j) + sz(j))ds\right)\cdot z(j)|\\ 
    &\leq \|\int_{0}^{1} D\varphi_k(x(j) + sz(j))ds\|_2 \|z(j)\|_2 \\ 
    &\leq N r^2 K_4 \xi.
\end{align*} \end{linenomath} 
So, this implies that $\|\Lambda(X, \bar{Z})\|_{\infty} \leq m N r^2 K_4 \xi$ and proves the lemma.
\end{proof}

\begin{proof}[Proof of Theorem \ref{thm:appendix_noise_reconstr}.2]
Using  \eqref{eq:Y_prime} and  \eqref{eq:phi_mu_Y} we have
\begin{linenomath} \begin{align*}
\bar{Y} &= \big(\Phi_{\nu_{\xi}}(Y) - \Lambda(X, \bar{Z})\big) C_{\nu_{\xi}} + \bar{Z} \\ 
&=  \Phi_{\nu_{\xi}}(Y) C_{\nu_{\xi}} + \bar{Z} - \Lambda(X, \bar{Z}) C_{\nu_{\xi}}. 
\end{align*} \end{linenomath} 
The above equation for each column $i \in [N]$ is given by an equation of the form $\bar{y} = \Phi_{\nu_{\xi}}(Y) c_{\nu_{\xi}} + \bar{u}_i$, where the perturbation is
\begin{linenomath} \begin{align*}
\bar{u} = \bar{z} - \Lambda(X, \bar{Z}) c_{\nu_{\xi}}.
\end{align*} \end{linenomath} 
Using Lemma \ref{lemma:appendix_change_of_data} the perturbation vector $\bar{u}$ is bounded as
\begin{linenomath} \begin{align*}
\|\bar{u}\|_2 \leq \sqrt{n} \xi + \sqrt{n} m N r^2 K_4 \xi \|c_{\nu_{\xi}}\|_{\infty} \\ 
= \sqrt{n} \xi \Big(1 + m N r^2 K_4 \|c_{\nu_{\xi}}\|_{\infty}\Big).
\end{align*} \end{linenomath} 
We apply Theorem \ref{thm:appendix_Candes_noise}, and this concludes the proof. 
\end{proof}

\section{Conclusions} 

In summary, we proposed a method to reconstruct sparse networks from noisy and limited data. Our approach blends ergodic theory of dynamical systems and compressive sensing to demonstrate that once a minimum length of time series is achieved, the EBP, particularly its extension QEBP, is a robust method to identify network structures from noisy data.  The main advantage of this method is that it enables to use of a smaller amount of time series (quadratically in the degree and log of the system size) as opposed to a linear dependence on the system size of the classical Basis Pursuit method.

We introduced the relaxing path algorithm that reconstructs the network as a weighted graph parametrized by the bound of the noise. Without prior knowledge of the statistical properties of the noise corrupting the data, this algorithm can reveal the network structure in an optimal interval of the tuned parameter.  Because a noisy and limited amount of length of time series arises typically in experimental settings, our findings apply to a wide range of chaotic systems.

\noindent
\textbf{Data and code availability.} All data necessary for the reproduction of the results, all simulations and analysis scripts are available in the Ergodic Basis Pursuit repository \cite{EBP_github}.

\noindent
\textbf{Acknowledgments.} We are indebted to Joseph Hart and Raj Roy for sharing the experimental data. We thank Arkady Pikovsky, Jeroen Lamb, Narcicegi Kiran, Zheng Bian for enlightening discussions. E.R.S. acknowledges support by FAPESP grant 2018/10349-4. This work was supported by the FAPESP CEMEAI Grant No. 2013/07375-0,  Serrapilheira Institute (Grant No.Serra-1709-16124) and Newton Advanced Fellow of the Royal Society NAF\textbackslash R1\textbackslash 180236).

\noindent
\textbf{Author contributions.} All authors designed and performed research; E.R.S. wrote the code and made the figures. E.R.S. and T.P. analyzed data; and all authors wrote the paper.	

\noindent
\textbf{Author declaration.}The authors declare no competing interest.

\section{Appendix}
\subsection*{Relaxing path algorithm for noisy data}
\label{sec:B_eps}  
\eqref{eq:ell_2_close_solution} quantifies the approximation accuracy w.r.t. to the sparse vector $c_{\nu}$. We can use it to estimate the entries' magnitude lying outside the support set of this sparse vector, $\mathcal{S} = \mathrm{supp}(c_{\nu})$. 
Let us denote $u_\mathcal{S}$ as the vector equal to $u$ on the index set $\mathcal{S}$ and zero on its complement $\mathcal{S}^c$. We can decompose $c_{\nu}^{\star}(\epsilon)$ into the sum of $c_{\nu, \mathcal{S}}^{\star}(\epsilon)$ and $c_{\nu, \mathcal{S}^c}^{\star}(\epsilon)$. Note that $\|c_{\nu, \mathcal{S}}^{\star}(\epsilon) - c_{\nu}\|_2^2 + \|c_{\nu, \mathcal{S}^c}^{\star}(\epsilon)\|_2^2 = \|c_{\nu}^{\star}(\epsilon) - c_{\nu}\|_2^2$ since $\mathcal{S}$ and $\mathcal{S}^c$ are disjoint, and it implies that 
$\|c_{\nu, \mathcal{S}^c}^{\star}(\epsilon)\|_2 \leq K_4 \epsilon.$ Hence, assuming the wrong entries are assigned at random, we consider that any entry of $c_{\nu}^{\star}(\epsilon)$ with a magnitude less than $\mathcal{O}(\epsilon/\sqrt{m})$ is zero.

Since the entries' magnitude supported in $\mathcal{S}^c$ are bounded by $K_2 \epsilon$, we discard the irrelevant connections (to represent the node dynamics) encoded in $c_{\nu}^{\star}(\epsilon)$ as we tune $\epsilon$. The idea is to tune $\epsilon$ and find the connections that are robust over different parameter values, the relevant connections. The challenge is that $\xi$ is unknown, as well as the other quantities that bound the error in  \eqref{eq:perturbed_noise_}. We look at this problem as a one-parameter family, searching the support set that persists over different $\epsilon$ and reconstructing the sparse network. We propose the relaxing path algorithm:

\begin{enumerate}
\item Select a set of equally spaced values $\epsilon_k$ within the interval $\mathcal{E} = [\varepsilon_{\min}, \varepsilon_{\max}]$. A pre-processing analysis can estimate the interval bounds \cite{cleveland90}.  
\item For each $\epsilon_k \in \mathcal{E}$ find the optimal solution to the  \eqref{eq:QBP_EBP}, the support $\mathcal{S}_k = \mathrm{supp}(c_{\nu}^{\star}(\epsilon_k))$ and $\mathcal{T}_k = \mathcal{S}_{k} \Delta \mathcal{S}_{k-1}$, where $\Delta$ corresponds to the symmetric difference of the two sets and checks the change in their cardinality \cite{Figueiredo_gradient_2007}. 
\item If $|\mathcal{T}_k| = 0$, the support has not changed, then stop, and the corresponding solution $c_{\nu}^{\star}(\epsilon_k)$ is returned. Otherwise, iterate $k \mapsto k + 1$ and repeat Step 2.
\end{enumerate} 

\section{References}
% Bibliography
\bibliographystyle{alpha}
\bibliography{reconstruct}

\newcommand{\etalchar}[1]{$^{#1}$}
\begin{thebibliography}{WHK{\etalchar{+}}18}

\bibitem[AVDB18]{agrawal2018rewriting}
Akshay Agrawal, Robin Verschueren, Steven Diamond, and Stephen Boyd.
\newblock A rewriting system for convex optimization problems.
\newblock {\em Journal of Control and Decision}, 5(1):42--60, 2018.

\bibitem[BPK16]{brunton2016discovering}
Steven~L. Brunton, Joshua~L. Proctor, and J.~Nathan Kutz.
\newblock Discovering governing equations from data by sparse identification of
  nonlinear dynamical systems.
\newblock {\em Proceedings of the National Academy of Sciences},
  113(15):3932--3937, 2016.

\bibitem[BWB{\etalchar{+}}09]{bohland2009}
Jason~W. Bohland, Caizhi Wu, Helen Barbas, Hemant Bokil, Mihail Bota, et~al.
\newblock {A Proposal for a Coordinated Effort for the Determination of
  Brainwide Neuroanatomical Connectivity in Model Organisms at a Mesoscopic
  Scale}.
\newblock {\em PLOS Computational Biology}, 5(3):1--9, 03 2009.

\bibitem[Can08]{CANDES_2008}
Emmanuel~J. Candès.
\newblock The restricted isometry property and its implications for compressed
  sensing.
\newblock {\em Comptes Rendus Mathematique}, 346(9):589 -- 592, 2008.

\bibitem[CCMT90]{cleveland90}
Robert~B. Cleveland, William~S. Cleveland, Jean~E. McRae, and Irma Terpenning.
\newblock Stl: A seasonal-trend decomposition procedure based on loess (with
  discussion).
\newblock {\em Journal of Official Statistics}, 6:3--73, 1990.

\bibitem[CGT14]{Chang2014}
Young~Hwan Chang, Joe~W. Gray, and Claire~J. Tomlin.
\newblock Exact reconstruction of gene regulatory networks using compressive
  sensing.
\newblock {\em BMC Bioinformatics}, 15(1):400, Dec 2014.

\bibitem[CL06]{FanChung}
Fan Chung and Linyuan Lu.
\newblock {\em Complex graphs and networks}, volume 107 of {\em CBMS Regional
  Conference Series in Mathematics}.
\newblock Published for the Conference Board of the Mathematical Sciences,
  Washington, DC; by the American Mathematical Society, Providence, RI, 2006.

\bibitem[CNHT17]{casadiego2017model}
Jose Casadiego, Mor Nitzan, Sarah Hallerberg, and Marc Timme.
\newblock Model-free inference of direct network interactions from nonlinear
  collective dynamics.
\newblock {\em Nature Communications}, 8(1):2192, 2017.

\bibitem[CT05]{Candes_2005_decoding}
E.J. Candes and T.~Tao.
\newblock Decoding by linear programming.
\newblock {\em IEEE Transactions on Information Theory}, 51(12):4203--4215,
  2005.

\bibitem[DB16]{diamond2016cvxpy}
Steven Diamond and Stephen Boyd.
\newblock {CVXPY}: {A} {P}ython-embedded modeling language for convex
  optimization.
\newblock {\em Journal of Machine Learning Research}, 17(83):1--5, 2016.

\bibitem[DCB13]{ECOS}
A.~Domahidi, E.~Chu, and S.~Boyd.
\newblock {ECOS}: {A}n {SOCP} solver for embedded systems.
\newblock In {\em European Control Conference (ECC)}, pages 3071--3076, 2013.

\bibitem[DET06]{Donoho_2006_mutual_coherence}
D.L. Donoho, M.~Elad, and V.N. Temlyakov.
\newblock Stable recovery of sparse overcomplete representations in the
  presence of noise.
\newblock {\em IEEE Transactions on Information Theory}, 52(1):6--18, 2006.

\bibitem[DH01]{Donoho_uncertanty_principle}
D.L. Donoho and X.~Huo.
\newblock Uncertainty principles and ideal atomic decomposition.
\newblock {\em IEEE Transactions on Information Theory}, 47(7):2845--2862,
  2001.

\bibitem[DX14]{dunkl_xu_2014}
Charles~F. Dunkl and Yuan Xu.
\newblock {\em Orthogonal Polynomials of Several Variables}.
\newblock Encyclopedia of Mathematics and its Applications. Cambridge
  University Press, 2 edition, 2014.

\bibitem[EBP]{EBP_github}
Code available at
  \url{https://github.com/edmilson-roque-santos/Ergodic-basis-pursuit}.

\bibitem[ETvSP20]{Eroglu_PRX_2020}
Deniz Eroglu, Matteo Tanzi, Sebastian van Strien, and Tiago Pereira.
\newblock Revealing dynamics, communities, and criticality from data.
\newblock {\em Physical Review X}, 10:021047, 2020.

\bibitem[FNW07]{Figueiredo_gradient_2007}
Mário A.~T. Figueiredo, Robert~D. Nowak, and Stephen~J. Wright.
\newblock Gradient projection for sparse reconstruction: Application to
  compressed sensing and other inverse problems.
\newblock {\em IEEE Journal of Selected Topics in Signal Processing},
  1(4):586--597, 2007.

\bibitem[FO14]{Ftorek_2014}
Branislav Ftorek and Pavol Or\H{s}ansky.
\newblock Korous type inequalities for orthogonal polynomials in two variables.
\newblock {\em Tatra Mountains Mathematical Publications}, 58(1):1--12, 2014.

\bibitem[Fol13]{folland2013real}
G.B. Folland.
\newblock {\em Real Analysis: Modern Techniques and Their Applications}.
\newblock Pure and Applied Mathematics: A Wiley Series of Texts, Monographs and
  Tracts. Wiley, 2013.

\bibitem[FR13]{Foucart_mathematical_compressive_sensing}
Simon Foucart and Holger Rauhut.
\newblock {\em A Mathematical Introduction to Compressive Sensing}.
\newblock Birkh{\"a}user Basel, 2013.

\bibitem[Guc14]{Guckenheimer_2014}
John Guckenheimer.
\newblock From data to dynamical systems.
\newblock {\em Nonlinearity}, 27(7):R41, jun 2014.

\bibitem[HS17]{Hang_2017_bernstein-type}
Hanyuan Hang and Ingo Steinwart.
\newblock {A Bernstein-type inequality for some mixing processes and dynamical
  systems with an application to learning}.
\newblock {\em The Annals of Statistics}, 45(2):708 -- 743, 2017.

\bibitem[HSWD15]{han2015robust}
Xiao Han, Zhesi Shen, Wen-Xu Wang, and Zengru Di.
\newblock Robust reconstruction of complex networks from sparse data.
\newblock {\em {Physical Review Letters}}, 114:028701, 2015.

\bibitem[HZRM19]{hart2019topological}
Joseph~D Hart, Yuanzhao Zhang, Rajarshi Roy, and Adilson~E Motter.
\newblock Topological control of synchronization patterns: Trading symmetry for
  stability.
\newblock {\em {Physical Review Letters}}, 122(5):058301, 2019.

\bibitem[Kur84]{Kuramoto84}
Y.~Kuramoto, editor.
\newblock {\em Chemical Oscillations, Waves and Turbulence}.
\newblock Dover, Berlin, 1984.

\bibitem[MBPK16]{mangan2016inferring}
Niall~M Mangan, Steven~L Brunton, Joshua~L Proctor, and J~Nathan Kutz.
\newblock Inferring biological networks by sparse identification of nonlinear
  dynamics.
\newblock {\em IEEE Transactions on Molecular, Biological and Multi-Scale
  Communications}, 2(1):52--63, 2016.

\bibitem[NRP21]{NOVAES_2021_basis_adaptation}
Marcel Novaes, Edmilson {Roque dos Santos}, and Tiago Pereira.
\newblock Recovering sparse networks: Basis adaptation and stability under
  extensions.
\newblock {\em Physica D: Nonlinear Phenomena}, 424:132895, 2021.

\bibitem[NS08]{Domenico_2008}
Domenico Napoletani and Timothy~D. Sauer.
\newblock Reconstructing the topology of sparsely connected dynamical networks.
\newblock {\em Phys. Rev. E}, 77:026103, 2008.

\bibitem[PvST20]{Tanzi_2020}
Tiago Pereira, Sebastian van Strien, and Matteo Tanzi.
\newblock Heterogeneously coupled maps: hub dynamics and emergence across
  connectivity layers.
\newblock {\em J. Eur. Math. Soc.}, 22:2183--2252, 2020.

\bibitem[PYGS16]{Pan_2016}
Wei Pan, Ye~Yuan, Jorge Gonçalves, and Guy-Bart Stan.
\newblock A sparse bayesian approach to the identification of nonlinear
  state-space systems.
\newblock {\em IEEE Transactions on Automatic Control}, 61(1):182--187, 2016.

\bibitem[Rac91]{rachev1991probability}
S.T. Rachev.
\newblock {\em Probability Metrics and the Stability of Stochastic Models}.
\newblock Wiley Series in Probability and Statistics - Applied Probability and
  Statistics Section. Wiley, 1991.

\bibitem[SPMS17]{RMP}
Tomislav Stankovski, Tiago Pereira, Peter~VE McClintock, and Aneta Stefanovska.
\newblock Coupling functions: universal insights into dynamical interaction
  mechanisms.
\newblock {\em Reviews of Modern Physics}, 89(4):045001, 2017.

\bibitem[STK05]{sporns2005human}
Olaf Sporns, Giulio Tononi, and Rolf K{\"o}tter.
\newblock The human connectome: a structural description of the human brain.
\newblock {\em PLoS Comput Biol}, 1(4):e42, 2005.

\bibitem[STW18]{Schaeffer_2018}
Hayden Schaeffer, Giang Tran, and Rachel Ward.
\newblock Extracting sparse high-dimensional dynamics from limited data.
\newblock {\em SIAM Journal on Applied Mathematics}, 78(6):3279--3295, 2018.

\bibitem[STWZ20]{Schaeffer_2020}
Hayden Schaeffer, Giang Tran, Rachel Ward, and Linan Zhang.
\newblock Extracting structured dynamical systems using sparse optimization
  with very few samples.
\newblock {\em Multiscale Modeling \& Simulation}, 18(4):1435--1461, 2020.

\bibitem[Sze39]{Szego_1939}
G\'abor Szeg\^o.
\newblock {\em Orthogonal polynomials}.
\newblock Colloquium publications (American Mathematical Society) v. 23.
  American mathematical society, New York city, 1939.

\bibitem[Tak06]{takens2006detecting}
Floris Takens.
\newblock Detecting strange attractors in turbulence.
\newblock In {\em Dynamical Systems and Turbulence, Warwick 1980: proceedings
  of a symposium held at the University of Warwick 1979/80}, pages 366--381.
  Springer, 2006.

\bibitem[TW17]{Tran_2017}
Giang Tran and Rachel Ward.
\newblock Exact recovery of chaotic systems from highly corrupted data.
\newblock {\em Multiscale Modeling \& Simulation}, 15(3):1108--1129, 2017.

\bibitem[WHK{\etalchar{+}}18]{wang2018}
Shuo Wang, Erik~D Herzog, Istv{\'a}n~Z Kiss, William~J Schwartz, Guy Bloch,
  Michael Sebek, Daniel Granados-Fuentes, Liang Wang, and Jr-Shin Li.
\newblock Inferring dynamic topology for decoding spatiotemporal structures in
  complex heterogeneous networks.
\newblock {\em Proceedings of the National Academy of Sciences},
  115(37):9300--9305, 2018.

\bibitem[Win01]{winfree2001}
Arthur~T Winfree.
\newblock {\em The geometry of biological time}, volume~12.
\newblock Springer Science \& Business Media, 2001.

\bibitem[WLG16]{wang2016}
Wen-Xu Wang, Ying-Cheng Lai, and Celso Grebogi.
\newblock Data based identification and prediction of nonlinear and complex
  dynamical systems.
\newblock {\em Physics Reports}, 644:1--76, 2016.

\bibitem[ZLZK11]{zamora2011}
Gorka Zamora-L{\'{o}}pez, Changsong Zhou, and J{\"{u}}rgen Kurths.
\newblock {Exploring brain function from anatomical connectivity}.
\newblock {\em Frontiers in Neuroscience}, 5(JUN):1--11, 2011.

\end{thebibliography}

%%%%%%%%%% Merge with supplemental materials %%%%%%%%%%
\clearpage
\setcounter{section}{0}
\renewcommand{\thesection}{S-\Roman{section}}

\addtocontents{toc}{\protect\setcounter{tocdepth}{0}}
%----------------------------- HEADER ----------------------------------
\title{Supplementary Information to \\ 
Robust reconstruction of sparse network dynamics}

%------------------------------------------------------------------

\section{Relaxing path algorithm}
\label{sec:relaxing_path}

\begin{figure}[h]
    \centering
    \includegraphics[width=1.0\textwidth]{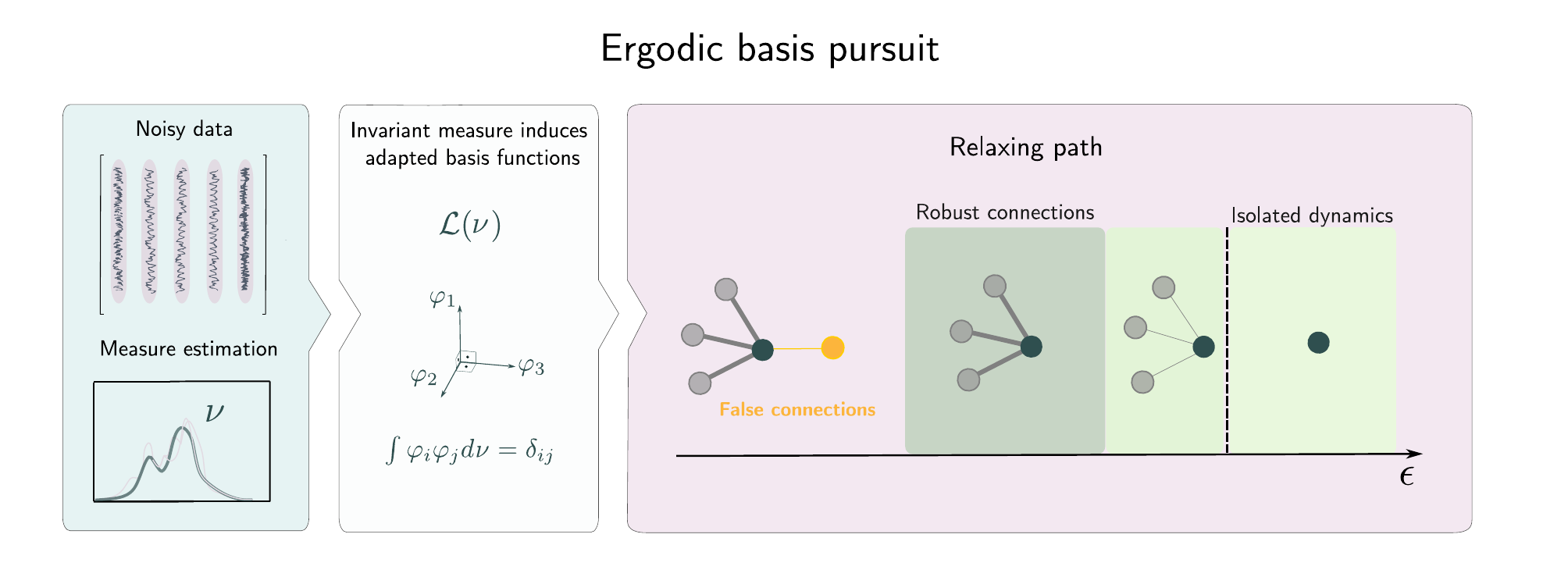}
    \caption{\textbf{Robust network reconstruction scheme using ergodic basis pursuit.} The noisy data is generated from a network dynamics whose underlying measure is $\mu_{\xi}$. Using its estimated measure $\nu$, we induce an orthonormal set of basis functions $\mathcal{L}(\nu)$ representing the dynamics. Under the assumption that the network dynamics is sparse, the noisy data and $\mathcal{L}(\nu)$ are recast as a minimization problem, whose solution encodes a proxy of the network. Although the noise level may be unknown, the relaxing path algorithm searches the connections of each node varying the noise level $\epsilon$ as a parameter. The true connections remain robust over an interval of $\epsilon$.}
    \label{fig:pipeline}
\end{figure}

Here we detail the relaxing path algorithm described in the Materials and Methods of the main text. This algorithm tunes the parameter $\epsilon$ to identify correct connections in the underlying network, see Figure \ref{fig:pipeline} for an illustration of the network reconstruction scheme using relaxing path algorithm.

Relaxing path algorithm uses the Theorem 7. To gain insight into what is happening, it is worth to consider a special case. Consider the case we probe a node in a weakly coupled network via the relaxing path algorithm. In this case, the true coefficient vector has the entries' magnitude relative to the isolated dynamics $f$ larger than those relative to the coupling function $h$. This difference also appears in the family of solutions $\{c_{\nu}^{\star}(\epsilon)\}_{\epsilon}$ as we increase $\epsilon$. For small $\epsilon$, we may observe a few false connections; see Figure \ref{fig:pipeline}. These false connections disappear for larger $\epsilon$, and only the robust connections remain. If we keep increasing $\epsilon$, the robust connections have their magnitude reduced until we only observe those entries relative to the isolated dynamics, i.e., there are no incoming connections. Thus, the parameter $\epsilon$ identifies the magnitude difference among the entries of the true solution, capturing the correct incoming connections to the node inside an appropriate interval of $\epsilon$.

Here, we detail the three main stages of the algorithm. 

\subsection{Model selection (MS)}
\label{sec:model_selection_}
Model selection selects the coefficient vector of a particular node for a given parameter value $\epsilon$. The current version of our algorithm encompasses the criterion used in the main text, where we search the robust connections in consecutive parameter values. Alternative approaches are valid and can be incorporated. For short we denote the model selection step as $MS(c^{\star}(\epsilon))$ for a given coefficient vector $c^{\star}(\epsilon)$. 

\subsection{Network selection}
\label{sec:net_selection}

Network selection is a map that obtains the network structure from the coefficient matrix. Although each coefficient matrix $C$ with column vectors $\{c_1, \dots, c_N\}$ is mapped to a directed multigraph (a graph that permits multiple edges among the nodes), the Network Selection introduces a map $NS$ that yields a graph structure instead. 

Let $\mathcal{S}_i \subset [m]$ be the set of indices corresponding to basis functions in $\mathcal{L}$ that depend on node $i$. Thus, the directed multigraph $G_m = ([N], E_m)$ from the representation of the network dynamics in $\mathcal{L}$ is defined by
\begin{align}\label{eq:multi_graph_edges}
E_m = \{(i, j) ~:~ w_{ij}^k = c_i^{k} \neq 0, k \in \mathcal{S}_j\},
\end{align}
where $(i, j)$ corresponds to an edge from node $j$ to $i$ and $w_{ij}^k$ is the edge weight. Self-loops are excluded since the graph only carries information about the coupling structure. The Network Selection step constructs a weighted graph from this multigraph, defined as
\begin{align}\label{eq:weighted_entries}
W_{ij} = \max_{k \in \mathcal{S}_j}\{w_{ij}^{k}\},
\end{align}
such that $W$ corresponds to an adjacency matrix whose entries are the edge weights in  \eqref{eq:weighted_entries}. We construct a directed subgraph of node $i$ picking the $i$-th row of $W$ and denote it by $\mathrm{subgraph}(c_i)$. The same construction can be replicated for all nodes, so the graph structure from the coefficient matrix $C$ is
\begin{align*}
G = \bigcup_{i = 1}^{N} \mathrm{subgraph}(c_i),
\end{align*}
characterizing the network selection map $NS: \mathbb{R}^{m \times N} \to \mathbb{R}^{N \times N}$ defined as $C \mapsto W$. When the edge weights are unnecessary, all edge weights are set to 1.  

\subsection{Algorithm}

The quadratically constrained Ergodic Basis Pursuit method is given in Algorithm \ref{alg:EBP_algorithm}. Let the hard-threshold function be given as
\begin{align*}
\mathrm{hard}(u, \lambda) = u \chi_{|u| > \lambda},
\end{align*}
where $\chi_{A}$ is the characteristic function on the set $A$. For a vector $u \in \mathbb{R}^{m}$, we consider that the hard-threshold function is evaluated coordinate-wise. The relaxing path algorithm reconstructs the networks as described in Algorithm \ref{alg:B_eps_algorithm}.

\begin{algorithm}[H]
 \SetAlgoLined
  \KwInput{$\bar{y} \in \mathbb{R}^{n}$, $\Phi_{\nu}(Y) \in \mathbb{R}^{n \times m}$, $\epsilon$\;}
    \tcc{$\bar{y} \in \mathbb{R}^{n}$: time series (measurement) vector}
	\tcc{$\Phi_{\nu}(Y) \in \mathbb{R}^{n \times m}$: adapted library matrix}
	\tcc{$\epsilon$: relaxing parameter}
  \KwOutput{$c^{\star}(\epsilon)$ coefficient vector\;}
	\textit{Output.} \begin{equation}\label{eq:QBP}
	    c_{\nu}^{\star}(\epsilon) = \argmin_{\tilde{u} \in \mathbb{R}^{m}} \{\|\tilde{u}\|_1 \quad \mathrm{subject~to}~ \|\Phi_{\nu}(Y) \tilde{u} - \bar{y}\|_2 \leq  \epsilon\}.
	\end{equation}
	\caption{Quadractically constrained Ergodic Basis Pursuit (QEBP)}
	\label{alg:EBP_algorithm}
\end{algorithm}

%%%%%%
\begin{algorithm}[H]
%\DontPrintSemicolon
\SetAlgoLined
  \KwInput{$\bar{Y} = (\bar{y}_1, \dots, \bar{y}_N) \in \mathbb{R}^{n \times N}$, $\Phi_{\nu}(Y) \in \mathbb{R}^{n \times m}$, $X \in \mathbb{R}^{m \times N}$\;} 
	\tcc{$\bar{Y}$: matrix of the noisy multivariate time series}
	\tcc{\texttt{criterion}: select criterion for Model Selection}
	\tcc{$\Phi_{\nu}(Y)$: adapted library matrix}
  \KwOutput{Reconstructed graph $G$\;}
	\textit{Initialization:} $G^0 = ([N], \emptyset)$, $\mathcal{F}^0 = \emptyset$, $\epsilon_{\min}, \epsilon_{\max} \in \mathbb{R}$ \; 
	\For{$j \in [N]$}
	{
	\For{$\epsilon \in [\epsilon_{\min}, \epsilon_{\max}]$}
	{
	
	$c_{\nu}^{\star}(\epsilon) = \mathrm{QEBP}(\bar{y}_j, \epsilon)$\;
	$c_{\nu}^{\star}(\epsilon) = \mathrm{hard}(c_{\nu}^{\star}(\epsilon), \epsilon/\sqrt{m})$\;
	$c_j^{\star}(\epsilon) = R_{\nu}^{-1} c_{\nu}^{\star}(\epsilon)$\;
	
	\If{MS$(c_j^{\star}(\epsilon), \texttt{criterion})$ is satisfied}{
	$H_j = \mathrm{subgraph}(c_j^{\star}(\epsilon))$ \;
	$G^{k + 1} = G^{k} \cup H_j$\;
	$\mathcal{F}^{k+1} = \mathcal{F}^k \cup \{j\}$\;
	\textbf{Stop} and go to next node\;
	}
	}
	}
	\textit{Output.} Coefficient matrix $C^{\star}$. 
\caption{Relaxing path algorithm}
 \label{alg:B_eps_algorithm}
\end{algorithm}

\section{Optoelectronic experimental data}
\label{sec:opto_electronic_details}
The experimental data corresponds to a network of optoelectronic oscillators whose nonlinear component is a Mach-Zehnder intensity modulator. Each node can be modeled as
\begin{align}\label{eq:appendix_opto_electronic_equation}
x_i(t+1) = \beta I_{\theta}(x_i(t)) +\alpha \sum_{j=1}^{17} A_{ij} [ I_{\theta}(x_j(t)) - I_{\theta}(x_i(t))] ~ \mbox{~mod~} 2\pi, \quad i = 1, \dots, N,
\end{align}
where the normalized intensity output of the Mach-Zehnder modulator is given by
\begin{align*}
I_{\theta}(x) = \sin^2(x + \theta),
\end{align*}
$x$ represents the normalized voltage applied to the modulator, $\beta$ is the feedback strength, and $\delta$ is the operating point set to $\frac{\pi}{4}$. We obtained the experimental multivariate time series $\{y_{1}(t), \cdots, y_{17}(t)\}_{t=1}^{16385}$ corresponding to $N = 17$ units coupled through a network, where the following parameters were fixed and known experimentally: $\beta = 4.5$ and coupling strength $\alpha = 0.171875$ \cite{hart2019topological}. Let us denote the optoelectronic network dynamics as $F$.

\subsection{Localization on a subset of the phase space}

\begin{figure}[h]
    \centering
    \includegraphics[width=1.0\textwidth]{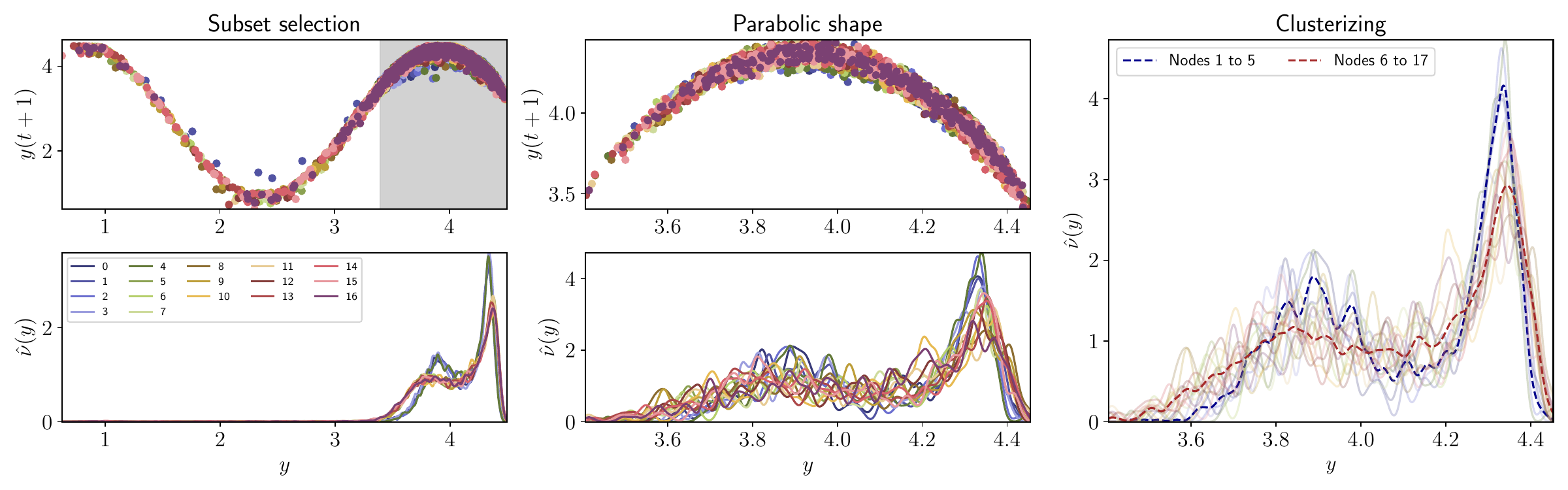}
    \caption{Step-by-step description of the extraction of the experimental data for coupling $\sigma = 0.171875$. The top middle panel displays the return map of the resulting network dynamics' trajectory with 264-time steps.}
    \label{fig:subset_selection}
\end{figure}

Figure \ref{fig:subset_selection} illustrates the extraction of experimental data used in the main text. 

\emph{Subset selection.} In Figure \ref{fig:subset_selection}, the upper left panel shows the return map for all nodes in the network measured experimentally (according to private communication, the initial $\sim$1000 points can be regarded as transient. Hence we removed it). The shaded gray area represents the phase space subset where nodes spend more time compared to other regions; in this case, it corresponds to the interval $\mathcal{I} = [3.4, 4.5]$. This is confirmed by the bottom left panel, which depicts the estimated density function for each node. All densities are concentrated inside the same gray area but decay rapidly to zero over the complement of the interval. 

\emph{Parabolic shape of the return map.} We extracted the most significant sample of the network dynamics' trajectory, satisfying that all nodes should remain inside the gray region. This subset selection yields a sample with a length of $224$ time steps and is shown in the top middle panel. The shape of the return map changes from a sinusoidal to a parabolic shape, suggesting using polynomial basis functions up to degree two in the library for the reconstruction procedure. The bottom middle panel shows the estimated density for the trajectories inside the subset. Observe that in view of the densities, either the bottom left and bottom middle panels, there are roughly two groups of nodes, represented by nodes 1 to 5 and 6 to 17 (near $y = 4.4$ this observation is clear). 

\emph{Clustering.} For each group of nodes, we consider all trajectories as different initial conditions of the same map and estimate the density function, as described in  \eqref{eq:density_estimation_1d}. In Figure~\ref{fig:subset_selection}, right panel depicts the density function  ${\color{blue} \rho_1 (y_1, \dots, y_5)}$ for group 1 to 5 in blue, and ${\color{red} \rho_2 (y_6, \dots, y_{17})}$ for group 6 to 17 in red. Note both density functions are smoother than the individual densities since we used more data to estimate them. The final estimated density we use to orthonormalize the polynomial basis functions is written as the product density function of both group density functions, i.e.,
\begin{align}\label{eq:product_blue_red}
    \rho(y_1, \dots, y_{17}) \approx {\color{blue} \rho_1 (y_1, \dots, y_5)} \times {\color{red} \rho_2 (y_6, \dots, y_{17})}.
\end{align}

\subsection{Basis functions selection preserves the network structure }

The parabolic shape of the return map in the top middle panel in Figure \ref{fig:subset_selection} corresponds to the restriction of the optoelectronic network dynamics $F$ onto $\mathcal{A}$, which we denote $\tilde{F}= F|_{\mathcal{A}}$. Hence, $\tilde{F}$ lies in the span of the quadratic polynomials. Taylor expanding \eqref{eq:appendix_opto_electronic_equation} at the maximum point inside the interval $\mathcal{A} = [3.4, 4.5]$, there are no crossed terms in the expansion. Hence, we only included monomials up to degree 2, so for each $j = 1, \dots, N$, the basis functions are $\phi_k(y_j) = y_j^k$ where $k = 1, 2$. The independent term $\phi_0(y) = 1$ is added with a single column in the library matrix. 

\subsection{Reconstruction of the network structure}
\label{sec:reconstr_opto}

To quantify the overall reconstruction performance, we introduce a weighted false link proportion for each node. Let $\mathcal{M}_i$ and $\hat{\mathcal{M}}_i$ be the subset of edges node $i$ shares with its neighbors of the original and estimated graph, respectively. We assume that any edge has a weight equal to 1, but those estimated using the relaxing algorithm and denoted by $\{w_{ij}\}_{i,j}$. So, we calculate the proportion of false positive (FP) and false negative (FN) at node $i$ as:
\begin{align*}
FP_i &= \frac{\sum_{j = 1}^N w_{ij} \chi_{\hat{\mathcal{M}}_i \cap \mathcal{M}_i^c}\big((i, j)\big)}{\sum_{j = 1}^N \Big( w_{ij} \chi_{\hat{\mathcal{M}}_i \cap \mathcal{M}_i^c}\big((i, j)\big) + \chi_{\hat{\mathcal{M}}_i^c \cap \mathcal{M}_i^c}\big((i, j)\big)\Big)}, \\ 
FN_i &= \frac{\sum_{j = 1}^N \chi_{\hat{\mathcal{M}}_i^c \cap \mathcal{M}_i}\big((i, j)\big)}{\sum_{j = 1}^N \chi_{\mathcal{M}_i}\big((i, j)\big)},
\end{align*}
where $\chi_{\mathcal{U}}$ is the indicator function of the subset $\mathcal{U}$. 

\section{Approximating invariant measures from multivariate time series}
\label{sec:invariant_measure}

In this section, we discuss how we estimate a proxy for the physical measure $\mu_{\alpha}$ from the multivariate time series $\{x(t)\}_{t = 0}^{n}$ (we write it for the noiseless case, but the same approach is used for the noisy measurements $\{y(t)\}_{t = 0}^{n}$).

Let $\{x(t)\}_{t = 0}^{n}$ be the multivariate time series on $M^N$ and consider the empirical measure $P_n = \frac{1}{n} \sum_{t = 0}^{n - 1} \delta_{x(t)}$ on $M^{N}$. To estimate the physical measure, we employ kernel density estimators. The kernel density estimator consists of taking the convolution (which we denote by $*$) of the empirical measure $P_n$ and the kernel $G_{\chi}: [0, \infty) \to [0, \infty)$ \cite{Hang_2018_KDEf}, where $\chi$ is the bandwidth parameter. We consider the Gaussian kernel throughout our results, i.e., $G_{\chi}(x) = e^{-x^2/\chi^2}$.

Hang and co-authors \cite{Hang_2018_KDEf} evaluated the convergence rate of kernel estimators for dynamical systems satisfying certain mixing conditions. Following there results, assume that $\mu_{\alpha}$ is absolutely continuous with respect to Lebesgue with density $\varrho_{\alpha}$. The rate of convergence of the estimated density with respect to $\varrho_{\alpha}$ in $L_1$-norm depends on the regularity of $\varrho_{\alpha}$. If $\varrho_{\alpha}$ is $\beta-$H\"older continuous then, it  behaves asymptotically for the length of time series as 
\begin{align}\label{eq:appendix_L_1_rate_of_convergence}
 \|\varrho_{\alpha} - G_{\chi} * P_n \|_1 \leq  \mathcal{O}\Big( \big(\frac{\log n^{3}}{n}\big)^{\frac{\beta}{2 \beta + N}}\Big).
\end{align}
Hence, in the case of large networks, the convergence is slow, which requires a large amount of time series. In \cite{Hang_2018_KDEf} the authors also consider more general regularity conditions, for instance pointwise $\beta$-H\"older controllable condition. However, there are no explicit expressions of the convergence rates. So, for sake of exposition we consider the case in \eqref{eq:appendix_L_1_rate_of_convergence}. 

\subsection{Product measure} Following our assumption in the paper, we assume that the physical measure $\mu_{\alpha}$ is close to a product measure $\nu$. So, instead of estimating $\mu_{\alpha}$, we estimate $\nu$ from the data. We consider the multivariate time series $\{x_i(t)\}_{t = 0, i = 1}^{n, N}$ as the observation from the same system, i.e., $\{x(t)\}_{t = 0}^{nN}$. More precisely, the estimation from data is formulated as follows: we consider the empirical measure $\frac{1}{nN} \sum_{t = 0}^{nN - 1} \delta_{x(t)}$. Hence, for a fixed $\chi > 0$, the proxy of $\varrho_{\alpha}$ is given by the convolution of the empirical measure and the kernel $G_{\chi}$
\begin{align}\label{eq:density_estimation_1d}
    \rho_{n, N, \chi}(x) = \frac{1}{n N \chi}\sum_{t = 0}^{nN - 1} G_{\chi}(x - x(t)).
\end{align}
In other words, the density of the product measure $\nu$ is given by $\prod_{i \in [N]}\rho_{n, N, \chi}$. Since the expression of $\rho_{n, N, \chi}$ corresponds to a sum of Gaussian kernels, it fulfils restrictions on the shape given in \cite{Hang_2018_KDEf}. In particular, we can deduce an upper bound of the Lipschitz constant of $\rho_{n, N, \chi}$ as follows:
\begin{lemma}
For a given $i \in [N]$ let $M_i = [a, b] \subset \mathbb{R}$ with $b > a$, and let $a_1 = \max\{|a|, |b|\}$. Then
\begin{align*}
\mathrm{Lip}(\prod_{i \in [N]}\rho_{n, N, \chi}) = \mathrm{Lip}(\rho_{n, N, \chi}) \leq \frac{4}{\chi^2} a_1.
\end{align*}
\end{lemma}
\begin{proof}
As before, here we consider $\mathrm{Lip}(\prod_{i \in [N]}\rho_{n, N, \chi}) = \max_{i \in [N]}\mathrm{Lip}(\rho_{n, N, \chi})$ then the first equality holds. By Mean-value theorem, the Lipschitz constant $\mathrm{Lip}(\rho_{n, N, \chi})$ is given by
\begin{align*}
\mathrm{Lip}(\rho_{n, N, \chi}) &= \|D\rho_{n, N, \chi}\|_{\infty} \\
&= \max_{x \in [a, b]} |D \rho_{n, N, \chi}(x)| \\
&= \max_{x \in [a, b]} \frac{2}{nN\chi^3} \left|\sum_{t = 0}^{nN - 1} (x - x(t)) e^{-\frac{(x - x(t))^2}{\chi^2}}\right| \\
&\leq  \frac{2}{\chi^2} ( \max\{|a|, |b|\} + \frac{1}{Nn} \sum_{t = 0}^{nN - 1} |x(t)| ) \\
&\leq\frac{4}{\chi^2} a_1.
\end{align*}
where we used that $\frac{1}{\chi}e^{-\frac{(x - x(t))^2}{\chi^2}} \leq 1$ for any $x \in [a, b]$, and the claim follows.
\end{proof}
We also guarantee that $\rho_{n, N, \chi}(x) > 0$ for any $x \in [a, b]$ then $\rho_{0} = \min_{i \in [N]} \{ \min_{x \in [a, b]} \rho_{n, N, \chi}(x)\} > 0$. More importantly, the $\|\prod_{i \in [N]}\rho_{n, N, \chi} - \varrho_{\alpha}\|_1$ has similar rate of convergence as ~\eqref{eq:appendix_L_1_rate_of_convergence}, but for the $d = 1$. So, the speed of convergence does not depend on the dimension of the phase space.

\end{document}